\documentclass[11pt]{article}
\newcommand{\version}{}

\usepackage{amsthm,amsfonts,amsmath}
\usepackage{
amssymb}
\usepackage{graphicx
,%pdf
}
\swapnumbers
\theoremstyle{plain}

\newtheorem{thm}{THEOREM}[section]
\newtheorem{lm}[thm]{LEMMA}

\newtheorem{remark}[thm]{Remark}

\theoremstyle{definition}

\theoremstyle{remark}
\newcommand{\upchi}{\raise1pt\hbox{$\chi$}}
\newcommand{\R}{{\mathord{\mathbb R}}}

\newcommand{\Z}{{\mathord{\mathbb Z}}}

\newcommand{\dd}{{\rm d}}

\newcommand{\F}{{\mathcal{F}}}

\numberwithin{equation}{section}
\let \e=\varepsilon                                                                
\begin{document}

%\iffalse
%\fi
%%%%%%%%%%%%%%%%%%%%%%
\markboth{\scriptsize{CCELM \version}}{\scriptsize{CCELM \version}}

\title{\bf{Droplet minimizers for the Gates-Lebowitz-Penrose free energy functional}}
\author{\vspace{5pt} E. A.  Carlen$^1$, M. C. Carvalho$^2$, R. Esposito$^3$, 
J. L. Lebowitz$^4$ and R. Marra$^5$\\}

\date{\version}
\maketitle

\footnotetext                                                                         
[1]{ Department of Mathematics, Rutgers University, 
Piscataway, NJ
08854, U.S.A. Work partially
supported by U.S. National Science Foundation
grant DMS 06-00037.  }
\footnotetext 
[2]{Department of Mathematics and CMAF, University of Lisbon,
1649-003 Lisbon, Portugal}
\footnotetext 
[3]{Dipartimento di
Matematica, Universit\`a dell'Aquila, Coppito, 67100 AQ, Italy}
\footnotetext 
[4]{Departments of Mathematics and
Physics, Rutgers University, New Brunswick, NJ
08903, U.S.A.}
\footnotetext 
[5]{Dipartimento di Fisica and Unit\`a INFN, Universit\`a di
Roma Tor Vergata, 00133 Roma, Italy. \\
\copyright\, 2009 by the authors. This paper may be reproduced, in its
entirety, for non-commercial purposes.}
                                                                      
\begin{abstract}

We study the structure of the constrained minimizers of the Gates-Lebowitz-Penrose free-energy functional ${\mathcal F}_{\rm GLP}(m)$,
non-local functional of a density field $m(x)$, $x\in {\mathcal T}_L$, a  $d$-dimensional 
torus  of side length $L$. At low temperatures,  ${\mathcal F}_{\rm GLP}$ is not convex, and has
two distinct global minimizers, corresponding to two equilibrium states. Here we constrain the average density $L^{-d}\int_{{\cal T}_L}m(x)\dd x$ to be a fixed value $n$ between the densities in the two equilibrium states, but close to the low density equilibrium value. 
In this case, a ``droplet'' of the high density phase may or may not form in a background of the low density phase, depending on the values $n$ and $L$.
We determine the
critical density for droplet formation, and the nature of the droplet, as a function of $n$ and $L$. 
\if false
Our free energy functional is obtained from the canonical microscopic free energy for a particle or spin system with an attractive long range Kac potential
via a  scaling limit.  The relation of the free energy to the large deviations functional for the mesoscopic density is proven in some cases, and expected to hold in general. 
\fi
The relation between the  free energy and the large deviations  functional for 
a particle model with long-range Kac potentials, proven in some cases, and expected to be true in general, then  provides information on the structure of  typical microscopic configurations of the Gibbs measure when the range of the Kac potential is large enough.
\end{abstract}

\bigskip
\centerline{Mathematics Subject Classification Numbers:  49S05, 52A40, 82B26}

\section{Introduction} \label{intro}

\subsection{The free energy functional} \label{intro1.1}

Let  ${\cal T}_L$ denote the $d$-dimensional torus with edge length $L$. Let 
${\cal M}$ denote the set of measurable functions $m$ from ${\cal T}_L$ to $[-1,1]$. Here, 
$m\in {\cal M}$ is an order parameter field representing the local average magnetization in an Ising system on a lattice
in ${\cal T}_L$, viewed on a  mesoscopic scale in which the  microscopic lattice structure is invisible. 
The underlying microscopic model has a well-known lattice gas interpretation, in which  $\rho := (1+m)/2$, which takes values in $[0,1]$, is viewed as a  particle density that is bounded above due to the microscopic constraint that at most one particle may occupy any lattice site.

The object of this paper is to investigate the minimizers, and near-minimizers, of certain mesoscopic free energy functionals on ${\cal M}$ under a constraint on the value of   the average magnetization
(or density)  $L^{-d}\int_{{\cal T}_L}m(x)\dd x$.  These free energy functionals  (defined just below) arose in work of
Lebowitz and Penrose \cite{LP}  and Gates and Penrose \cite{GP},  
and have been proven to be large deviations functionals
for the underlying Ising systems, at least in certain cases \cite{ABCP,P}.  Thus ``near minimal free energy profiles'' $m$ correspond to ``typical'' coarse grained magnetization configurations for the Gibbs measure of the underlying Ising system.

We investigate the nature of the (near-)minimal free energy profiles $m$ for sub-critical temperatures, so that there are two
equilibrium values $m_\pm$ of the magnetization, and  for values of $n$ just slightly above $m_-$. 
Depending on the amount of the excess magnetization $n - m_-$, the surface tension between the two phases, the bulk compressibility, and on $L$, the
excess magnetization will either be uniformly dispersed over the entire volume, or will partly aggregate into a ``droplet'' of the $m_+$ phase in a sea of the $m_-$ phase. 
This phenomena was studied rigorously on the microscopic scale for the two dimensional Ising model with nearest neighbor interactions by Biskup, Chayes and Kotecky \cite{BCKb}. 

Here we consider this same droplet formation problem, but in any dimension $d$, and for Ising systems with long range Kac potentials instead of nearest neighbor interactions. While  \cite{BCKb} presents  direct analysis of the microscopic system, our starting point is instead to analyze a variational problem for  a mesoscopic
free energy functional where a number of tools from the calculus of variations can be used. 
%This has certain advantages. 
%After all, the droplet itself, when it exists, will be a (nearly) ?? macroscopic object, and thus our starting point is closer to the scale on which the droplet lives. 
Moreover, the connection between the microscopic Gibbs measure and the mesoscopic free energy functional has already been thoroughly investigated
\cite{ABCP,P}, and we can rely on this to make contact with the microscopic scale. Finally, the mesoscopic variational problem is of interest in its own right.

The Gates-Lebowitz-Penrose (GLP) free energy functional $\F_{\rm GLP}$ that we study has the form
\begin{equation}\label{GLP}
\F_{\rm GLP}(m) := \int_{{\cal T}_L}f(m(x))\dd x - 
\frac{1}{2}\int_{{\cal T}_L}\int_{{\cal T}_L}m(x)J(x-y)m(y)\dd x\dd y\ ,
\end{equation}
where $f$ is some convex function, and $J$ is a non-negative radial function with
\begin{equation}\label{normal}
\int_{\R^d}J(x)\dd x =1\ .
\end{equation}
We further suppose that $J$ has finite range range, which we  then use to set our length scale so that $J$ has unit range. 
More specifically, we require $J(x) = 0$ for $|x|>1$, and $J(x)> a$, some $a>0$,  for $|x| \le 1/2$.

Such functionals  were introduced by Gates and Penrose \cite{GP}, building on previous work by Lebowitz and Penrose \cite{LP}.  The function $f$ is a local free energy function taking into account short range interactions and entropy effects. 
The term involving $J$ is the interaction energy  due to  a long-range, local mean field type interaction among the spins (particles), mediated by an interaction potential $-\gamma^d J(\gamma r)$ where $\gamma^{-1}$ is the range of the potential in units of the microscopic lattice spacing, and $r$ is a lattice coordinate. 
This corresponds to the spin Hamiltonian 
\begin{equation}\label{range}
H = -\gamma^d\sum_{x,y \in \Z^d \bigcap \gamma^{-1}{\cal T}_L } J(\gamma |x-y|)\sigma(x)\sigma(y)\ 
\end{equation}
which specifies the long term attractive interaction.

The book \cite{P} by Presutti may be consulted for more information on the relation between this functional and the underlying Ising system.   Full details may be found there  for the case in which the only short range interaction is a hard-core repulsion corresponding in the particle picture  to the restriction that there is at most one particle at each lattice  site. Then,
with $\beta$ denoting the inverse temperature, the local free energy $f(m)$ is simply the $-\beta$
times the lattice gas  entropy term
\begin{equation}\label{ent}
s(m)=-\frac{1-m}{2}\log\frac{1-m}{2}-\frac{1+m}{2}\log\frac{1+m}{2} \ ;
\end{equation}
see \cite{ABCP,P}.

We shall focus on the case  $f =-\beta^{-1} s$ here as well, though our analysis is readily adapted to the
more general case of a strictly convex local free energy functional $f(m)$; see \cite{BP}
\if false
However, in the interest of simplicity of exposition and of making as close a contact as possible with the literature on the microscopic model, we 
postpone discussion of the general case, and first explain the issues and results for this special choice of 
$f(m)$.
\fi

The mesoscopic length scale is set by the range of the $J$, i.e., $\gamma^{-1}$ lattice spacings.
The macroscopic length scale  is $L$, the size of the domain. In the scaling limit that leads to
$\F_{\rm GLP}$ as a large deviations functional, the ratio between $L$ and the range of $J$ is held fixed
(so that $L$ is also measured in units of $\gamma^{-1}$, like the range of $J$), while $\gamma$ tends to zero.  This is very different from the frequently encountered scalings in which one also 
takes a thermodynamic limit ($L\to \infty$) at the same time one takes the continuum limit ($\gamma\to 0$), in which case
 the size $L$ of the system grows faster rate than $\gamma^{-1}$; e.g., $\gamma^{-2}$.

Our problem is to determine the minimum value of $\F_{\rm GLP}$ subject to the constraint
\begin{equation}\label{const}
\frac{1}{L^d}\int_{{\cal T}_K}m(x)\dd x = n\ ,
\end{equation}
and to characterize all of the  profiles $m$ that nearly minimize ${\cal F}_{\rm GLP}$.

The nature of  this minimization problem is clarified through use of the identity
$$-\int_{{\cal T}_L}\int_{{\cal T}_L}m(x)J(x-y)m(y)\dd x\dd y = 
\frac{1}{2}\int_{{\cal T}_L}\int_{{\cal T}_L}[m(x)-m(y)]^2|J(x-y) {\rm d}x{\rm d}y - \int_{{\cal T}_L}m^2(x)\dd x\ ,
$$
which is valid for $L>1$ (which we always take to be the case) on account of (\ref{normal}), 
to write the free energy functional in the form
\begin{equation}\label{GLP2}
{\cal F}_{\rm GLP}(m) =\int_{{\cal T}_L}\left[ f(m(x)) -\frac{1}{2}m^2(x)\right] {\rm d}x  + 
\frac{1}{4}\int_{{\cal T}_L}\int_{{\cal T}_L}[m(x)-m(y)]^2J(x-y) {\rm d}x{\rm d}y\ .
\end{equation}

In case $f(m) - m^2/2$ is a convex function of $m$, the minimization problem is trivial: One subtracts an appropriate multiple $\mu m$ from $f(m)$ so that the modified local free energy is minimized at $m=n$. 
Because of the constraint (\ref{const}),
this modification has no effect on the minimizers. Then, it is clear that the uniform profile $m(x) = n$ is the unique minimizer.  However,  $f(m) - m^2/2$ need not be convex. 

For example, with $f(m) = -s(m)/\beta$, the function $f(m) - m^2/2$
is strictly convex if $\beta \le 1$, but for $\beta>1$, it is not. Instead, it is a 
``double well'' function with minima at $\pm m_\beta$, where $m_\beta$ is the positive solution to
$m_\beta = {\rm tanh}(\beta m_\beta)$.
In this case, $\pm m_\beta$ are the two equilibrium values of $m$, $m_\pm$,  mentioned earlier. 
To simplify the writing of $\F$ for this case on which we shall focus, let us introduce the function $F$ on $[-1,1]$
defined by
$$F(m) =  \left[ -\frac{1}{\beta}s(m) -\frac{1}{2}m^2\right] -  \left[ -\frac{1}{\beta}s(m_\beta) -\frac{1}{2}m_\beta^2\right] \ ;$$
this differs from $-s(m)/\beta - m^2/2$ by a constant so that $F(m) \ge 0$, and $F(m) = 0$ if and only if
$m = \pm \beta$. 

We then define $\F$ to be 
\begin{equation}\label{GP}
{\cal F}(m) :=\int_{{\cal T}_L}F(m(x)) {\rm d}x  + 
\frac{1}{4}\int_{{\cal T}_L}\int_{{\cal T}_L}[m(x)-m(y)]^2J(x-y) {\rm d}x{\rm d}y\ .
\end{equation}

The problem of determining
${\displaystyle \inf_{m\in {\cal M}} \left\{ {\cal F}(m)\ :\ \frac{1}{L^d}\int_{{\cal T}_L}m(x){\rm d}x =n\ \right\}}$ 
and finding all  minimizing profiles $m$  is very easy for $n\in [-1,-m_\beta]$ and $n\in [m_\beta, 1]$: In these intervals, the unique minimizer is the constant profile $m(x) = n$.  (One easy way to see this is pointed out in Remark~\ref{trivial}.) 
Hence, the question is only interesting for $n\in (-m_\beta,m_\beta)$.

This question has been well studied for {\em  fixed} values of $n \in (-m_\beta,m_\beta)$ as $L$ tends to infinity (see \cite{P} and references quoted therein). Here
we are concerned with values of $n$ such that
$$n = -m_\beta+{\cal O}(L^{-\frac d{d+1}})\ ,$$
which turns out to be critical for droplet formation: If the value of $n$ is low enough in this range, the minimizing profiles will be uniform, and if it is large enough in this range, the minimizing profile
will represent a ``droplet'' of the $+m_\beta$ phase in a sea of the $-m_\beta$ phase; i.e., the droplet is the region in which $m(x) \approx m_\beta$.

In the well-studied problem, with fixed $n\in (-m_\beta,m_\beta)$ and sufficiently large $L$, one always sees a droplet  which has the {\em equimolar volume} $D_0$,
 specified by
\begin{equation}\label{eqi}
m_\beta D_0 - m_\beta  (L^d - D_0) = n L^d\qquad{\rm and\ hence}\qquad D_0 = \frac{n+m_\beta}{2m_\beta}L^d\ .
\end{equation}Indeed, if a profile $m$ takes on only the values $\pm m_\beta$ then the constraint (\ref{const})
will be satisfied if and only if the volume of the set on which it has the value $+m_\beta$ is $D_0$.

We shall see that when $m_\beta+n = {\cal O}(L^{-\frac{d} {d+1}})$ and $L$ is large, then the droplets,
when they exist, are smaller than this,
but not too small:  The volume $D$ of the droplet will satisfy
$$\left(\frac{2}{d+1}\right)D_0 \ \le\  D \ \le\  D_0\ .$$
That is, there is a {\it universal} lower bound on the size of stable droplets. 
Droplets that are ``too small'' always ``prefer to evaporate''.  The factor $\displaystyle{\frac{2}{d+1}}$ is independent of the particular interaction potential $J$,
and if one considers $n(L)$ defined by $n(L) = -m_\beta + KL^{-\frac{d}{d+1}}$, one can determine the precise
fraction of the equimolar volume as a function of $K$. 
\medskip

Results of this type were first obtained for the $2$-dimensional nearest neighbor Ising model by Biskup, Chayes and Kotecky \cite{BCK,BCKb}.  In  \cite{BCK}, they presented a general heuristic analysis of droplet formation,
which predicts the above universal lower bound on $D$,  and then they proved in \cite{BCKb} that the predictions of the heuristic analysis were correct for the nearest neighbor  Ising model, by a very detailed analysis of
the microscopic states that support  the Gibbs measure.  This rigorous analysis was carried out directly on the microscopic level, and did not involve the analysis of a free energy functional. However, their heuristic analysis, which was based on a competition between surface tension and compressibility, could well be expected to apply to the 
GLP functional, and other free energy functionals like it.

This expectation has been borne out 
in recent work of ourselves \cite{CCELM1}, and also of Belletini, Gelli, Luckhaus, Novaga \cite{BGLN}.
In these papers,  the problem of determining the nature of the minimizing profiles has been solved  for
a phenomenological analog of the GLP free energy functional, the
Allen--Cahn free energy functional: 
\begin{equation}\label{AC}
{\cal F}_{\rm (AC)}(m) = 
 \frac{1}{4}\int_{{\cal T}_L} (1-m^2(x))^2{\rm d}x +
  \frac{\theta^2}{2}\int_{{\cal T}_L}|\nabla m(x)|^2{\rm d}x\ .
\end{equation}
Here $(1-m^2)^2/4$ is a simple caricature of the more physical  double well potential $F(m)$, and
the gradient integral penalizes variation in $m$ just as 
\begin{equation}\label{nonloc}
\frac{1}{4}\int_{{\cal T}_L}\int_{{\cal T}_L}[m(x)-m(y)]^2|J(x-y) {\rm d}x {\rm d}y 
\end{equation}
does.  For purposes of this paper, we may regard  ${\cal F}_{\rm (AC)}$ as a 
phenomenological caricature of GLP functional.
Though ${\cal F}_{\rm (AC)}$ is often called the Allen--Cahn, Cahn-Hilliard or Landau--Ginzburg
functional,  it goes back to van der Waals \cite{db,VW}. 

The analysis of  ${\cal F}_{\rm (AC)}$ is much easier than the analysis of $\F$ because many tools that are applicable to the local interaction term $\int_{{\cal T}_L}|\nabla m(x)|^2{\rm d}x$ cannot be applied to its non-local relative 
(\ref{nonloc}).  In particular, both papers \cite{CCELM1} and \cite{BGLN} made use of an 
an idea originating with Modica \cite{M} that used the  {\em co-area formula} \cite{Fed} 
to get lower bounds on ${\cal F}_{\rm (AC)}(m)$.
The co-area formula
is simply the change of variables in which one uses $m$ itself as one of the variables of integration:
$${\rm d}^dx = \frac{1}{|\nabla m(x)|}{\rm d}\sigma_h {\rm d}h\ ,$$
where  ${\rm d}\sigma_h$ is the $(d-1)$-dimensional Huasdorff measure on  the level surface $\Gamma_h= \{x\ :\ m(x) = h\}$.
Using this, 
$$ {\cal F}_{\rm (AC)}(m) = \int_\R \int_{\Gamma_h}\left(
\frac{\theta^2}{2}|\nabla m(x)| +  \frac{1}{4}  \frac{(1-h^2)^2}{|\nabla m(x)|}\right) {\rm d}\sigma_h{\rm d}h\ .$$
Then, by the arithmetic--geometric mean inequality, we have the lower bound
\begin{eqnarray}\label{coco}
{\cal F}_{\rm (AC)}(m) &\ge&  \int_\R \int_{\Gamma_h}
\left(
\frac{\theta}{2}|1-h^2|
\right) {\rm d}\sigma_h{\rm d}h\nonumber\\
&\ge&   \int_{[-1+\e,1-\e]}
\frac{\theta}{2}(1-h^2) |\Gamma_h| {\rm d}h\ ,\nonumber\\
\end{eqnarray}
where $\e>0$, and $|\Gamma_h|$ is the $(d-1)$-dimensional Huassdorf measure of $\Gamma_h$.  
Since for each $h$ in the retained domain of integration, $\Gamma_h$ bounds a set including 
the set $D_\e = \{ x \,:\, \ m(x) > 1-\e\}$, the isoperimetic inequality gives us a lower bound on
$|\Gamma_h|$ in terms of the $d$-dimensional volume $|D_\e|$.

There is more to be done to determine the nature of the minimizers; see \cite{CCELM1,BGLN}.
But  the co-area formula, first used in this context by Modica and Mortola \cite{MM}, provides a key to the analysis of the Allen--Cahn functional that does not seem to be adaptable to the non-local context where the interaction  is not given in terms of a gradient. 

On a technical level, our main innovations here are the development of tools to obtain a sharp lower bound in the non-local case. We shall make essential use of various rearrangement inequalities, combined with various truncation arguments and {\em a priori} estimates on near minimizers. The technical difficulties are  worthwhile, we believe, because the GLP free energy functional has a direct connection with an underlying microscopic model, unlike the Allen-Cahn free energy functional.

The rest of the paper is organized as follows. We first briefly recall the heuristic analysis in \cite{BCK}. From this we deduce some natural conjectures about the nature of the minimizers of (\ref{GP}). The conjectures are correct, so we state them as theorems, and prove them in the next sections. 
We conclude with a short section discussing further properties of the minimizers, and some open questions concerning them.

\section{The heuristic analysis}
\medskip

Consider a free energy functional $\F_{\rm GLP}$ as in (\ref{GLP2}) for some strictly convex function $f(m)$ and some $\beta>0$.  Let $g(m) = 
f(m) -  \beta m^2/2$, and suppose that this is a ``double well potenetial'' so that there is a tangent line
$am+b$ that touches the graph of $g(m)$ exactly twice. Let $(m_-,g(m_-))$ and $(m_+,g(m_+)$, $m_- < m_+$,  denote these two
points. Then the functions $g(m) - (am+b)$ is strictly positive except at $m = m_\pm$, where it is zero.
Notice that for any profile $m(x)$ satisiying (\ref{const}), $\int_{{\cal T}_L}(am(x) + b)\dd x = L^d(an + b)$, and so
subtracting $am +b$ from $g(m)$ simply subtracts a fixed constant from 
\begin{equation}\label{GLP3}
\inf_{m\in {\cal M}} \left\{ {\cal F}_{\rm GLP}(m)\ :\ \frac{1}{L^d}\int_{{\cal T}_L}m(x){\rm d}x =n\ \right\}\ ,
\end{equation}
and has no effect on the class of minimizers.

Consider the case in which the constraint value $n$  in (\ref{GLP3}) is {\em slightly} above $m_-$, and well below $m_+$. There are several obvious options to consider when trying to construct profiles $m(x)$ that will yield minimal, or nearly minimal, values in (\ref{GLP3}).

One option might be to put all of the excess of $m$ over $m_-$  into a ``droplet'' in which $m(x) = m_+$, and outside of which $m(x) = m_-$.   The constraint requires the volume  of the droplet; i.e., the region in which $m(x) \approx m_+$, would
to be the equimolar volume $D_0$ given
by $m_+ D_0 - m_-  (L^d - D_0) = n L^d$.

However, the optimal  shape of the droplet
would depend on the symmetry properties of  $J$.  In general, for  $L$ large compared to $D_0^{1/d}$, one would expect  the
cost of forming a droplet to come from the {\em surface tension}  between the  $m_-$ and $m_+$  regions; see \cite{RWi}.   The shape that minimizes the surface tension is known as the {\em Wulff shape} for the functional 
${\cal F}_{\rm GLP}$.  Whatever the Wulff shape turns out to be,  one would expect that for   $L$ large compared to $D_0^{1/d}$,
 the free energy cost of forming a droplet of volume $D_0$ should be proportional to 
$D_0^{1- 1/d}$.

In particular, under our assumption that  $J$ is isotropic, we would expect the optimal droplet to be very nearly a ball of volume $D_0$, and the free energy cost of forming the droplet to be simply a multiple $S$, the {\em surface tension}, of
the surface area of a ball of volume $D_0$.

Therefore, letting  $\sigma_d$ denote the surface area of the unit sphere in $\R^d$, we would have
\begin{equation}\label{dren}
{\cal F}(m) \approx   S\sigma_d\left(\frac{D_0}{\sigma_d/d}\right)^{1-1/d} \ 
\end{equation}
for a profile that arranges the excess magnetization into a round droplet of volume $D_0$.  Without the assumption that 
$J$ is isotropic, the proportionality constant would be different, depending on the nature of the Wulff shape, but
the cost of this option would still be proportional to  $D_0^{1- 1/d}$. 

Another option might be to smear the excess uniformly over the background. This gives a bulk contribution,
and its size is determined by the compressibility $\chi_-$ which is the inverse of $g''(m_-)$, where, as above, $g(m) = f(m) - \beta m^2/2$.
Smearing a droplet of volume $D_0$  over ${\cal T}_L$, we get the uniform profile
\begin{equation}\label{nfo}
m(x) =  n = m_- + (m_+ - m_-)\frac{D_0}{L^d}\ .
\end{equation}
  Hence, when $L^d$ is large compared to $D_0$,
$$g(m(x)) \approx g(m_-) + \frac{1}{2}g''(m_-)\left[(m_+- m_-)\frac{D_0}{L^{d}}\right]^2$$
for all $x$. Integrating over the domain, we find that this option for $m$ gives
$${\cal F}_{\rm GLP}(m)  \approx  \frac{(m_+- m_-)^2}{2\chi_-}\frac{D_0^2}{L^d}\ .$$

Which option does better?  The free energy in the droplet option is independent of $L$, while the cost of smearing the droplet over the backgound decreases as $L$ increases. So with $D_0$ held fixed, 
the droplet does better for small values of $L$, and the uniform profile does better for large values of 
$L$. To determine the break even point, equate the two values for ${\cal F}$ to find
$$S\sigma_d\left(\frac{D_0}{\sigma_d/d}\right)^{1-\frac 1d} =   \frac{(m_+- m_-)^2}{2\chi_-}\frac{D_0^2}{L^d}\ .$$
Thus, both are comparable when $D_0^{1+\frac 1d} = {\cal O}(L^d)$, which by (\ref{nfo}) means
\begin{equation}\label{critreg}
n +m_-= {\cal O}(L^{-\frac d{d+1}})\ .
\end{equation}
This defines the {\em critical scaling regime}.

What should one expect for the minimizing free energy  in the critical scaling regime, and will the minimizers
be given by some sort of droplet, or not?

In \cite{BCK}, 
Biskup, Chayes and Kotecky proposed that to answer this question, 
one should introduce a {\it volume fraction} $0 \le \eta \le 1$, and put $\eta D_0$ into a drop of the appropriate (Wulff) shape, and $(1-\eta)D_0$  into the
uniform background.  They then constructed a phenomenological thermodynamic free energy function $\Phi(\eta)$
 which is the sum of the surface tension term and the uniform background term. In the case that $J$ is isotropic, so that the Wulff shape is a ball, $\Phi(\eta)$ is given by
$$
\Phi(\eta) =  S\sigma_d\left(\frac{\eta D_0}{\sigma_d/d}\right)^{1-1/d} + 
\frac{(m_+- m_-)^2}{2\chi_-}\frac{((1 - \eta)D_0)^2}{ L^d}
$$

The suggestion of \cite{BCK}  is that in great generality, one can resolve a  competition
between surface and bulk energy  effects by choosing $\eta\in [0,1]$ to minimize
$\Phi$. This picture was then proven rigorously in \cite{BCKb} for the $2$-dimensional Ising model
with nearest neighbor interactions, where the Wulff shape is temperature dependent, and droplets correspond to connetcted clusters of $+$ spins in a sea of $-$ spins.

\subsection{The  reduced variational problem}
\medskip

Define   $C(D_0,L)$ by
\begin{equation}\label{CCd}
C(D_0,L) := \frac{(m_+- m_-)^2}{2d \chi_- S} \left(\frac{d}{\sigma_d}\right)^{1/d}
\frac{D_0^{1+1/d}}{L^d}\ .
\end{equation}
Then  $\Phi(\eta)$ can be written as
$$\Phi(\eta) =  S\sigma_d\left(\frac{ D_0}{\sigma_d/d}\right)^{1-1/d}\left[\eta^{1-1/d} + C(D_0,L)(1 - \eta)^2\right]\ .$$

For fixed $D_0$ and $L$, the function $\Phi(\eta)$ is  increasing at $\eta = 0$ (with an infinite derivative),  and has a local minimum for some $\eta>0$. Depending on the value of $C(D_0,L)$, the absolute minimum of $\Phi$ on $[0,1]$ may be at either $\eta = 0$, or at the local minimum at positive $\eta$. In the first case,
there is no droplet; everything gets smeared over the background. In the latter case, one has a droplet
with a  radius corresponding to the fraction of the excess that one puts in the droplet. 

To solve  this minimization problem, 
notice that $\Phi(\eta) - \Phi(0)$ is a constant multiple of
\begin{equation}\label{phidiff} \eta[ \eta^{-\frac 1d} + C(D_0,L)\eta - 2C(D_0,L)]\ ,
\end{equation}
which vanishes at $\eta = 0$, and hence  has a minimum at some $\eta>0$ if and only if it  becomes negative somewhere. For which values of $D_0$ and $L$ does this happen?
By the arithmetic--geometric mean inequality,
\begin{eqnarray}
 \eta^{-\frac 1d} + C\eta &=& \frac{d}{d+1}\left(\frac{d+1}{ d}\eta^{-\frac 1d}\right) +
 \frac{1}{d+1}\Big((d+1)C\eta\Big)\nonumber\\
 &\ge&   \left(\frac{d+1}{d}\eta^{-\frac 1d}\right)^{\frac d{d+1}}\Big((d+1)C\eta\Big)^{\frac 1{d+1}}\\
 &=&  C^{\frac 1{d+1}}\frac{d+1}{d^{\frac d{d+1}}}\ .\nonumber
\end{eqnarray}
 Thus, the quantity in (\ref{phidiff}) is minimized  at $\eta > 0$ if and only if 
 $$C(D_0,L)^{\frac1{d+1}}\frac{d+1}{ d^{\frac d{d+1}}} \le 2C(D_0,L)\ .$$
Let $C_\star$ be the value of $C$ that gives equality in this last inequality. One finds, 
$$
C_\star = \frac{1}{d}\left(\frac{d+1}{2}\right)^{\frac{d+1}{d}}\ .
$$
Moreover, with $C(D_0,L) = C_\star$, there is equality in the application made above 
of the arithmetic geometric mean inequality if and only if $\displaystyle{\frac {d+1}{d}\eta^{-\frac 1d}} = (d+1)C_\star\eta$.  
Therefore, define $\eta_\star$ by $\eta_\star = (d C_\star)^{-\frac d {d+1}}$. One finds
\begin{equation}\label{est}
\eta_\star = \frac 2{d+1}\ .
\end{equation}

That is, if the minimum of $\Phi$ is attained at some positive value of $\eta$, the positive value
is never less than $\eta_\star$.  The volume of a droplet will always lie between that of the
equimolar droplet $D_0$, and the reduced value  $\eta_\star D_0$. Smaller droplets are never seen; they prefer to evaporate.  
Thus, the simple ansatz of dividing the excess volume between a droplet and
the background, and optimizing over the volume fraction yields  simple predictions for whether one sees a droplet of not, and the size of the droplet if there is one.  

Of course, the shape of the droplet was put into the ansatz by hand. The surface tension formula
${\displaystyle S\sigma_d\left(\frac{ D_0}{\sigma_d/d}\right)^{1-1/d}}$ is what one expects to be
appropriate  for the isotropic long range  interaction 
that we consider here. For a non-isotropic interaction, one should  have to replace this surface tension formula
by the corresponding formula for the appropriate Wulff shape. The crucial point is that this would still be some multiple of
$D_0^{1-1/d}$, and would therefore lead in the same way  to the analysis of 
$\eta[ \eta^{-\frac 1d} + C(D_0,L)\eta - 2C(D_0,L)]$, except for a different constant $C(D_0,L)$.

\if false
That this volume fraction optimization ansatz does indeed yield the correct description of ``what one sees'' 
 was rigorously established for the  $2$ dimensional Ising model in \cite{BCKb}.  In our previous work  \cite{CCELM}, and in \cite{BGLN}, this was rigorously proved for the
Allen--Cahn free energy functional. We shall now prove that it is also correct for the GLP free energy functional. 
\fi

\subsection{Statements of the theorems on the minimizers}
\medskip

The theorems we present in this section refer to the specific GLP free energy functional $\F$ given in (\ref{GP}).
This model has been particularly well studied in the literature (see \cite{P} and the references there).
As we have explained above,  for $n$ close to $-m_\beta$ the free energy of a round droplet of equimolar 
volume $D_0$ should be given by (\ref{dren}),
and the ansatz of \cite{BCK} suggests that

\begin{equation}\label{GLPV}
\inf_{m\in {\cal M}} \left\{ {\cal F}(m)\ :\ \frac{1}{L^d}\int_{{\cal T}_L}m(x){\rm d}x =n\ \right\} \approx
\inf_{\eta\in [0.1]}S\sigma_d\left(\frac{ D_0}{\sigma_d/d}\right)^{1-1/d}\left[\eta^{1-1/d} + C(D_0,L)(1 - \eta)^2\right]\ ,
\end{equation}
where $C(D_0,L)$ is given by (\ref{CCd}) using (\ref{eqi}) to express $D_0$ in terms of $n$ and $L$, and using the
values of $\chi_-$ and $S$ appropriate to this particular form of $\F$. Of course, $\chi_-$ can be easily computed from the explicit form of $f(m) = -(1/\beta) s(m) - m^2/2$ given in (\ref{ent}). In fact, since $s(m)$, and hence
$f(m)$ is symmetric in $m$, the compressibility is the same in the two phases, and we shall simply write $\chi$ in 
place of $\chi_-$ for this model.

As for the surface tension,  an explicit variational formula for $S$
is given in  (\ref{pst}) in the next section. Suffice it to say here that it is the minimal value for a simpler variational problem concerning  one dimension for profiles interpolating between $-m_\beta$ and $+m_\beta$. 

We shall investigate (\ref{GLPV}) in the critical scaling regime (\ref{critreg}), for which $n+m_\beta$ is proportional to $L^{-\frac{d}{d+1}}$.
Therefore, fix any $K>0$, and define
$$  n =  -m_\beta + KL^{-\frac{d}{d+1}}\ .$$
With this choice of $n$, (\ref{eqi}) gives us
\begin{equation}\label{elim}
D_0 = \frac{K}{2m_\beta}L^{\frac{d^2}{d+1}} \quad{\rm and}\qquad 
S\sigma_d\left(\frac{ D_0}{\sigma_d/d}\right)^{1-1/d} = 
S\sigma_d\left(\frac{ Kd}{2m_\beta \sigma_d}\right)^{1-\frac{1}{d}}L^{\frac{d^2-d}{d+1}}\ .
\end{equation}

Inserting this value of $D_0$ in (\ref{CCd}), one finds that $C(D_0,L) = C(K)$, a constant depending only on $K$
that is given by
\begin{equation}\label{dkd}
C(K)  := 
\frac{2m_\beta^2}{d\chi S}\left(\frac{d}{\sigma_d}\right)^{\frac{1}{d}}
\left(\frac{K}{2 m_\beta}\right)^{1+\frac{1}{d}}\ .
\end{equation}

Therefore, with $n =  -m_\beta + KL^{-\frac{d}{d+1}}$,  (\ref{GLPV}) becomes
\begin{multline}
\inf_{m\in {\cal M}} \left\{ {\cal F}(m)\ :\ \frac{1}{L^d}\int_{{\cal T}_L}m(x){\rm d}x =n\ \right\} \approx\nonumber\\
L^{\frac{d^2-d}{d+1}} S\sigma_d\left(\frac{ Kd}{2m_\beta \sigma_d}\right)^{1-\frac{1}{d}}
\inf_{\eta\in [0.1]}\left[\eta^{1-1/d} + C(K)(1 - \eta)^2\right]\ ,
\end{multline}

Our first theorem says that if we divide both sides through by $L^{\frac{d^2-d}{d+1}} $, this becomes exact in the limit as $L$ tends to infinity:

\begin{thm}\label{thm1}
 For all $K>0$, 
\begin{multline}\label{limone}
\lim_{L\to \infty} L^{-\frac{d^2-d}{d+1}} {{\displaystyle \inf_{m\in {\cal M}} \left\{ {\cal F}(m)\ :\ \frac{1}{L^d}\int_{{\cal T}_L}m(x){\rm d}x =
 -m_\beta + KL^{-\frac{d}{d+1}}\ \right\}}}{}
 = \\
S\sigma_d\left(\frac{ Kd}{2m_\beta \sigma_d}\right)^{1-\frac{1}{d}}
\inf_{\eta\in [0.1]}\left[\eta^{1-1/d} + C(K)(1 - \eta)^2\right]\ ,
\end{multline}
where
$C(K)$ is given by (\ref{dkd}).
\end{thm}

By what has been explained in the previous section, the infimum on the right in  (\ref{limone}) occurs if and only if 
$$C(K)  > C_\star =  \frac{1}{d}\left(\frac{d+1}{2}\right)^{\frac{d+1}{d}}\ .$$
Evidently from (\ref{dkd}), this is the case if and only if $K > K_*$ where
\begin{equation}\label{kstd}
K_\star = \frac{d+1}{2} m_\beta \left(\frac{\chi  S}{2m_\beta^2}\right)^{\frac{d}{d+1}}\left(\frac{\sigma_d}{d}\right)^{\frac {1}{d+1}}
\ .
\end{equation}

\medskip

Theorem~\ref{thm1} therefore suggests that the curve 
$$n_c(L) = -m_\beta + K_\star L^{-\frac d{d+1}}$$ is critical for droplet formation,
so that for large $L$ and densities $n$ significantly below this level, the minimizers will be uniform, while
for large $L$ and densities $n$ significantly above this level, the minimizers will correspond to spherical droplets
of a reduced volume  $\eta_c D_0$ where $\eta_c$ is given by (\ref{est}).   The following theorems bear this out.

\begin{thm}\label{thm2} 
For all $K < K_\star$ and $L$ sufficiently large, when
$$-1 \le n \le -m_\beta + KL^{-\frac d{d+1}}\ ,$$
the unique minimizer for ${\cal F}$ is the uniform order parameter field $m(x) = n$. 
\end{thm}

To show that droplets do form for $K > K_\star$, we need a precise definition of
we mean by a ``droplet of the $+m_\beta$ state in a sea of the $-m_\beta$ state''. 
Toward this end, set $\kappa=\big(KL^{-\frac{d}{d+1}}\big)^{\frac 1 3}$ and define the subsets  $A$ and $C$ 
of ${{\cal T}_L}$ by slicing ${{\cal T}_L}$ at the following level curves of $m$:
$$A = \{\ x\in {{\cal T}_L}\ :\ -m_\beta+\kappa \le m(x) \le m_\beta-\kappa\ \}\ ,\quad C =  \{\ x\in {{\cal T}_L}\ :\  m(x) \ge m_\beta-\kappa\ \}. $$

\begin{thm}\label{thm3} 
Suppose $K > K_\star$, and
$n = -m_\beta + KL^{-\frac d{d+1}}$.
Then, given $\e>0$  one can find $\alpha>0$ so that for any trial function $m$ for which ${\cal F}(m)<f_L(n)+\alpha$ the following statements are true:
$$\big||C\cup A|-\eta_cD_0\big|<\e D_0$$
and
$$\big||C|-\eta_cD_0\big|<\e D_0\ ,$$
where $\eta_c$ is the optimal volume fraction from Theorem~\ref{thm1}. 
\end{thm}

\begin{remark} Since $m(x)$ is close to, or larger than, $m_\beta$ on $C$, and is
close to, or smaller than, $-m_\beta$ on ${\cal T}_L\backslash(A\cup C)$, we may think of $C$ as the ``droplet of condensate'' and $A$ as the (evidently thin, and presumably anular) surface region of the droplet.  The estimates of
Theorem~\ref{thm3} specify only the size of the droplet, and not its shape. The analysis of this shape is the subject of continuing research. It is heuristically clear from Theorem~\ref{thm1}
that this optimal shape is very close to a ball. Further discussion on what is required to prove this may be found in the final section.
\end{remark}

\medskip
To prove the theorems we need good upper and lower bounds for ${\cal F}(m)$ at admissible trial functions $m$.
The upper bounds come from a trial function suggested by the ansatz in \cite{BCK}.  The lower bound is the part that is more technically challenging for the reasons explained above. We begin with the upper bound.

\section{The upper bound}

\subsection{A good trial function for the intermediate regime}

The arguments of \cite{BCK} suggest that one should use as a
trial function a function of the form
$$
m_{(\eta)}(x) =  m_0(|x| - \eta^{\frac 1 d}r_0) +  \alpha(\eta)\ ,
$$
where:
\smallskip
\begin{enumerate}
\item $r_0$ is the radius of a ball of volume $D_0$ (known as the {\em equimolar radius}), and so
$ \eta^{\frac 1 d}r_0$ is the radius of a ball with volume $\eta D_0$.
\item $m_0$ is some one dimensional transition profile that very nearly minimizes the cost in free energy of making
the transition from $m = -m_\beta$ to $+m_\beta$ around the origin. 
\item  $\alpha(\eta)$ is a constant determined by the constraint ${\displaystyle \int_{{\cal T}_L} m_{(\eta)}(x){\rm d} x = nL^d}$.
\smallskip 
\end{enumerate}
Do not confuse $m_{(0)}$ which is a functions on ${\cal T}_L$, with $m_0$, which is a function on $\R$. 

\noindent{$\bullet$} {\it 
As $\eta$ varies in the interval  $0 < \eta < 1$, this family of ``fractional droplet'' trial functions
{\em interpolates} between smearing everything over the background, for $\eta=0$, and    putting everything into the equimolar droplet, for $\eta =1$.}

\subsection{Planar surface tension and the choice of $m_0$}

The natural choice for $m_0$ is given by minimizing the transition cost:
\begin{equation}
S  = \inf\left\{\ \int_{\R}  \left[ [f(m(z)) - f(m_\beta)]+\frac{1}{ 4}\int_{\R}|m(z) - m(y)|^2\bar J(z-y){\rm d}y \right]\dd z\ :\ 
\lim_{z\to \pm \infty}m(z) = \mp m_\beta\ \right\}\ \label{pst}
\end{equation}
where, writing $x\in \R^d$ as $x = (y,z)$ with $y\in \R^{d-1}$ and $z\in \R$,
\begin{equation}
\bar J(z) = \int_{\R^{d-1}} J(y,z){\rm d}y\ \label{jbar}
\end{equation}
and $J$ is the interaction potential in the GLP free energy functional. 

The quantity $S$  is  the {\em planar surface tension}; see \cite{P}. It is well known \cite{DOPT, P} that the minimizer $\bar m$ is unique up to translations. In the rest of the paper, $\bar m$ is the minimizer vanishing at the origin.

We  now choose $m_0$. We cannot simply choose $m_0 = \bar m$ since then
$m_0(|x| - \eta^{\frac 1 d}r_0)$ would not define a smooth, or even continuous,  function on ${{\cal T}_L}$.

However, only mild modifications are required:
We modify it so that  $m_0(|x| - \eta^{1/d}r_0)$ defines  a smooth function on ${{\cal T}_L}$, and the difference
between $m_0$ and $\bar m$ goes to zero exponentially fast as $L$ tends to infinity. We define $m_0(z)$ as any smooth function on $\R$ such that 
$$m_0(z) =  
\begin{cases} 
\phantom{-} \bar m(z)  &{\rm  if}\quad  |z| < L^{\frac{d-1}{d+1}} \\ -m_\beta{\rm sgn}(z)  
&{\rm  if}\quad   |z| > 2L^{\frac{d-1}{d+1}}  \\
\end{cases}\ .$$
By (\ref{elim}),  in the critical scaling regime $r_0$ is proportional to $L^{\frac{d}{d+1}}$, so that while
$L^{\frac{d-1}{d+1}}$ is large, it is small compared to $r_0$.

\subsection{The determination of $\alpha(\eta)$}

The constraint equation is
${\displaystyle \int_{{\cal T}_L} m_{(\eta)}(x){\rm d}x = nL^d}$,
and since  (by definition of $r_0$) 
$$nL^d =    m_\beta\frac{\sigma_d} d r^d_0 -m_\beta\left(L^d- \frac{\sigma_d} d r^d_0\right)\ ,$$
we have
\begin{equation}\label{alphaform}
\alpha(\eta) =   \frac1 { L^d}\left[2m_\beta\frac{\sigma_d} d r^d_0 -L^d\right]  - 
\frac1 { L^d}\int_{{\cal T}_L} m_0(|x| - r_\eta){\rm d}x\ .
\end{equation}
We require sharp estimates on the integral on the right. But we know enough about $m_0$ to derive them, and can now estimate ${\cal F}(m_{(\eta)})$ quite closely:

\medskip

\begin{lm}\label{uplem}  In the critical scaling regime, with $n = -m_\beta + KL^{-\frac{d}{d+1}}$,  
$$
{\cal F}(m_{(\eta)})  \le  L^{\frac{d^2-d}{d+1}} S\sigma_d\left(\frac{ Kd}{2m_\beta \sigma_d}\right)^{1-\frac{1}{d}}
\left[\eta^{1-1/d} + C(K)(1 - \eta)^2\right]    +
{\cal O}\left(L^{\frac{d^2-2d}{d+1}} \right)\ .$$
\end{lm}

Note that  $L^{\frac{d^2-2d}{d+1}}$ proportional to  $r_0^{d-2}$.
In proving Lemma \ref{uplem}, it will be convenient to express our estimates in terms of powers of $r_0$
instead of powers of $L$. To prove Lemma \ref{uplem} we start with:

\begin{lm} \label{intbnd}  For all $\eta$ such that $L^{\frac{d-1}{d+1}} < r_0\eta^{\frac{1}{d}} < r_0$,  
$$
\int_{{\cal T}_L} m_0(|x| - r_0\eta^{\frac 1 d}){\rm d}x  = m_\beta\left[2\frac{\sigma_d}{d}r_0^d \eta  -L^d\right]  
+{\cal O}(r_0^{d-2})
$$
so that
$$ 
\alpha(\eta)  = 2m_\beta \frac{\sigma_d}{d}
\frac{r_0^d}{L^d} (1-\eta) + {\cal O}\left(\frac{r_0^{d-2}}{L^d}\right)\ .$$
\end{lm}

\medskip

\noindent{\bf Proof:} Define $p(z)=-m_\beta{\rm sgn}(z)$ and the constant $M$ by
$M =  \int_\R z\big(p(z) - \bar m(z)\big){\rm d}z$,
and set $r_\eta=r_0\eta^{\frac 1 d}$.  Note that
$$\int_{{\cal T}_L} m_0(|x| - r_\eta){\rm d}x = m_\beta \left[2\frac{\sigma_d}{d}r_\eta^d -L^d\right]    - 
\int_{{\cal T}_L} (p(|x| - r_\eta) - m_0(|x| - r_\eta)) {\rm d}x\ .$$
Define
$$I_1 =  \int_{|x| \le 2r_\eta}p(|x| - r_\eta) - m_0(|x| - r_\eta)) {\rm d}x,\quad \text{ and } \quad I_2 =  \int_{|x| > 2r_\eta}(p(|x| - r_\eta) - m_0(|x| - r_\eta)) {\rm d}x\ .$$
We easily see that for all dimensions $d$,
$I_2 = {\cal O}(e^{-L^{1/4}})$.
Moreover, using polar coordinates,
$$I_1 = \sigma_d \int_0^{2r_\eta}\big(p(s - r_\eta) - m_0(s - r_\eta)\big) s^{d-1}{\rm d}s\ .$$
Introducing the new variable $z =s-r_\eta$, we see that if we extend the integration in $z$ over the whole real line, we only make an error of size ${\cal O}(e^{-L^{1/4}})$ at most, and so
$$I_1 =  \sigma_d r_\eta^{d-1}\int_\R (p(z) - m_0(z)) \left(1 + \frac{z}{ r_\eta}\right)^{d-1}{\rm d}z +
{\cal O}(e^{-L^{1/4}})\ .$$
Taking into account the fact that $(p(z) - m_0(z))$ is odd and rapidly decaying, we see that for $d=2$ or $d=3$,
$$\int_\R (p(z) - m_0(z)) \left(1 + \frac{z}{ r_\eta}\right)^{d-1}{\rm d}z = \frac{d-1}{ r_\eta}
\int_\R (p(z) - m_0(z))z{\rm d}z\ .$$
In higher dimension this  gives the leading order correction. 
This, together with the definition of $m_0(z)$
in terms of $\bar m(z)$,  yields the bound on the integral. Then the bound on $\alpha(\eta)$ follows from this and  (\ref{alphaform}).\qed
\medskip

\begin{remark}
We see from Lemma~\ref{intbnd} that in the critical scaling regime, except when $\eta =1$,
\begin{equation}\label{a2}
\alpha(\eta) \asymp L^{-\frac d{d+1}}\ .
\end{equation}
\end{remark}

\subsection{Computation of ${\cal F}(m_{(\eta)})$} \label{ub4}

With the trial function specified, we are now ready to prove Lemma~\ref{uplem}

\medskip

\noindent{\bf Proof of Lemma~\ref{uplem}:}   To simplify the notation, we write $m_0$ to denote $m_0(|x| - r_\eta)$ and $\alpha$
to denote $\alpha(\eta)$ 
so that $m_{(\eta)} = m_0 + \alpha$. We begin by estimating $\int_{{\cal T}_L}F((m_{(\eta)})\dd x$. Making a Taylor expansion, we find that 
for some $\lambda\in [0,1]$,
$$F(m_{(\eta)}) = F(m_0) + F'(m_0)\alpha + \frac{1}{ 2}F''(m_0)\alpha^2+\frac{1}{3!} F'''(m_0+\lambda \alpha)\alpha^3 .$$

We are required to produce a close upper bound on the integral of each of these terms over $\Omega$.
It turns out that the terms with odd derivatives are negligible, and that
to a very good of approximation
\begin{eqnarray}\nonumber
\int_{{\cal T}_L}F(m_{(\eta)})\dd x  &\approx& 
\int_{{\cal T}_L}F(m_0)\dd x + \frac{1}{2}\int_{{\cal T}_L}F''(m_0)\dd x \alpha^2\label{twoterms}\\
&\approx&
\int_{{\cal T}_L}F(m_0)\dd x + \frac{1}{2\chi} L^d \alpha^2\ .\nonumber\\
\end{eqnarray}

To see why this should be so, before going into the detailed calculations, note that $F(-m_\beta) = F(m_\beta) = 0$, and $m_0$ is essentially equal
to $\pm m_\beta$ except in a shell of unit thickness and radius $r_0\eta^{\frac{1}{d}}$. Thus, the term $\int_{{\cal T}_L}F(m_0)\dd x$ is ${\cal O}(r_0^{d-1})$.  Likewise, since  $F'(-m_\beta) = F'(m_\beta) = 0$, the integral of 
$F'(m_0)$ over ${\cal T}_L$ is ${\cal O}(r_0^{d-1})$. However, this integral gets multiplied by $\alpha$,
which is small. Hence this term is negligible compared to the first term. 

When we come to the second derivative term, we have ${\displaystyle F''(-m_\beta) = F''(m_\beta) = \frac{1}{\chi}}$
and so the integral of  $F''(m_0)$ over ${\cal T}_L$ is very close to ${\displaystyle \frac{L^d}{\chi}}$.  Since this integral
gets multiplied by $\alpha^2$, and we have $L^d\alpha^2\asymp L^dL^{-\frac {2d}{d+1}}={\cal O}(r_0^{d-1})$, this contribution it is of the same order as the first integral in the critical scaling regime.

Likewise, we have an ${\cal O}(L^d)$ bound on the integral of  $F'''(m_0+\lambda \alpha)$ over 
${\cal T}_L$, but while  the integral involving $F''$ gets multiplied by  $\alpha^2$, this integral gets
multiplied by $\alpha^3$, and so it too is negligible compared to the two integrals we shall keep.
The next several paragraphs contain the precise calculations, and then we turn to the interaction term.

Note that $F'(m) = \displaystyle{\frac 1 2\log\frac{1+m}{1-m} -m}$ and $F'(\pm m_\beta)=0$. Since  $F'(\bar m(z))$ is an odd, rapidly decaying function of $z$, estimates just like the ones employed in the proof of Lemma (\ref{intbnd})
show that 
$$\int_{{\cal T}_L}  F'(m_0) {\rm d} x = \sigma_dr_\eta^{d-1}\int_\R F'(\bar m(z) )
\left(1 + \frac{z}{ r_\eta}\right)^{d-1}{\rm d}z + {\cal O}(e^{-L^{1/4}})\ .$$
Then, with the constant $B$ defined by
${\displaystyle B =  \int_\R F'(\bar m(z))z{\rm d}z}$,
we have for $d=2$ or $d=3$ that
\begin{equation}\label{fp1}
\int_{{\cal T}_L}  F'(m_0) {\rm d} x = \sigma_d r_0^{d-2}B \eta^{\frac{d-2}d} + {\cal O}(e^{-L^{1/4}})\ ,
\end{equation}
and in any dimension $d\ge 4$, the term $ \sigma_d r_0^{d-2}B \eta^{\frac{d-2}d}$ gives the leading correction.

\noindent Next, $F''(m_0) = \displaystyle{\frac{1}{\beta(1-m_0^2)}-1 = \frac1{\chi(m_0)}}$.  Therefore, with $\chi=\chi(m_\beta)$,  
\begin{equation}\label{fp2}
\int_{{\cal T}_L}\frac{1}{ 2}F''(m_0){\rm d}x  = \frac{1}{ 2\chi(m_\beta)} L^d + R, \quad |R|\le \left(\frac 1{\chi}-\frac 1 {\chi(0)}\right)(1-\eta)\frac{\sigma_d} d r_0^d\ .
\end{equation}
Therefore the second order contribution is
$$\frac {1}{\sigma_d r_0^{d-1}}\alpha^2 \int_{{\cal T}_L}\frac{1}{ 2}F''(m_0){\rm d}x=\frac{2m_\beta^2}{\chi}\frac{\sigma_d}d(1-\eta))^2+{\cal O}(L^{-\frac d {d+1}})$$

\noindent Finally, $F'''(m) = \displaystyle{\frac{2 m}{(1-m^2)^2}}$, and so, for suffiently large $L$,
\begin{equation}\label{fp3}
\frac {1}{\sigma_d r_0^{d-1}}\alpha^3\left|\int_{{\cal T}_L}\frac{1}{3!} F'''(m_0+\lambda\alpha){\rm d}x \right|\le \frac {1}{\sigma_d r_0^{d-1}}\frac 2 \beta L^d\alpha^3={\cal O}(L^{-\frac{d}{d+1}}).
\end{equation}

We now combine these estimates with an estimate on the interaction term. First note that by Lemma~\ref{intbnd},  the
term ${\displaystyle  \frac{1}{2\chi} L^d \alpha^2}$ in (\ref{twoterms}) is exactly the bulk term in $\Phi_0(\eta)$.
We shall combine the other term, $\int_{{\cal T}_L}F(m_0)\dd x$, with the interaction terms, yielding:
$$\int_{{\cal T}_L}  F(m_0){\rm d}x +\frac{1}{4}
 \int_{{\cal T}_L}  \int_{{\cal T}_L}  J(x-y)[m_0(x)-m_0(y)]^2  {\rm d}x  {\rm d}y\ .$$
It remains to extract the surface contribution to  $\Phi_0(\eta)$ from these terms. 

We shall use the following simple fact: 
 For any function $g(y)$ depending only on $|y|$, and $|x| > 1$, the range of $J$, we have
\begin{equation}\label{curve}
\int_{\R^2}J(x-y)g(|y|) \dd y  = \left(1+ {\cal O}\left(\frac{1}{|x|-1}\right)\right)
\int_0 ^\infty  \overline J (|x|-s)g(s)ds\ .
 \end{equation}
A proof of such a statement in an even more general setting may be found in \cite{GL}. To prove the statement
we need here, we simply change integration variables $y \mapsto (|y|, z(y))$, $z(y)\in \R^{d-1}$ in such a way that
$|y-x|^2 = (|y| - |x|)^2 + z(y)^2$.  The change of variables doing this is explicitly given as follows:
For $s\in \R^+$ and $z\in \R^{d-1}$, set $u=\frac {x}{|x|}$ and
$$y= \frac{s\left(u+ \displaystyle{\frac{z}{|z|}}\gamma(s,z)\right)}{\sqrt{1+\gamma(s,z)^2}}\ , \quad \quad
\gamma(s,z)= \sqrt{\frac{(2|x|\/s)^2}{(2|x|\/s-|z|^2)^2}-1}\ .$$
Clearly $|y|=s$ and it is easy to check that $|x-y|^2-(|x|-|y|)^2=|z|^2$. Moreover, 
$$\frac{\gamma(s,z)}{|z|} = {\cal O}\left(\frac 1{|x|}\right),$$
uniformly in $y$ in the unit ball around $x$, provided that $|x|$ is sufficiently large. 

Since $J$ has unit range, we only need to bound the Jacobian of this transformation in the unit ball about $x$, and it is easy to see that for $|x| > 1$ this Jacobian differs from unity by an amount that is uniformly bounded in the unit ball about $x$ by a multiple of $(|x| -1)^{-1}$.

Once more, estimates just
like the ones employed in the proof of Lemma \ref{intbnd}
show that for some constant $c$,
\begin{eqnarray*}&&\int_{{\cal T}_L} {\rm d}x \left[F(m_0(x)+\frac{1}{4}\int_{{\cal T}_L} {\rm d}y J(x-y)[m_0(x)-m_0(y)]^2\right ]\approx\\
&&\sigma_dr_\eta^{d-1} (1 + \frac c{r_\eta})\int_\R {\rm d}z\left[F(\bar m(z))+\int_\R {\rm d}z' \bar J(z-z')[\bar m(z)-\bar m(z')]^2\right],
\end{eqnarray*}
where the errors are exponentially small in $L^{1/4}$. But because $m_0$ is so close to $\bar m$,
this only differs from  $S\sigma_d r_0^{d-1}\eta^{1-1/d}$ by errors that are ${\cal O}(r_0^{d-2})$.  
In the asymptotic scaling regime, $r_0^{d-2} \asymp L^{\frac{d^2-2d}{d+1}}$.

Combining the estimates, we have the proof of the lemma \qed

\section{The lower bound}
\medskip

The idea is, as in \cite{CCELMa}, to separate the surface and bulk contributions. The bulk estimate is similar to the one in \cite{CCELMa}, while the surface estimate requires new ideas based on rearrangement arguments. The key to the lower bound is a  {\it partition of ${{\cal T}_L}$} 
into three pieces:
\medskip
\begin{enumerate}
\item A region which will contribute a surface tension term to the free energy, 
\item A region which will contribute a compressibility term, 
\item A region that will make a negligible contribution.
\end{enumerate}
\medskip

To do this we fix a number $\kappa>0$ to be determined below.
Define numbers $h_+$ and $h_-$ by
\begin{equation}\label{hpmdef}
h_+ = m_\beta -\kappa\qquad{\rm and}\qquad h_- = -m_\beta + \kappa\ .
\end{equation}
Define the sets $A$, $B$ and $C$ by slicing ${{\cal T}_L}$ at the corresponding level curves:
\begin{eqnarray}
 A &=& \{\ x\in {{\cal T}_L}\ :\ h_- \le m(x) \le h_+\ \}\nonumber\\ 
 B &=&  \{\ x\in {{\cal T}_L}\ :\ m(x) \le h_-\ \}\nonumber \\
C &=&  \{\ x\in {{\cal T}_L}\ :\  m(x) \ge h_+\ \}. \label{ABCdef}
\end{eqnarray}
We denote by $I_A$, $I_B$ and $I_C$ the contribution to ${\cal F}(m)$ from the sets $A$, $B$ and $C$.

Define a radius $R$ by
\begin{equation}\label{Rdef}
\frac{\sigma_d}d R^d =  |C|\ ,
\end{equation}
where the right hand side denotes the measure of $C$.   Evidently $R$ is the radius of the ball with 
the same volume as $C$. 

It will be convenient in this section to  write $n$ in the form 
\begin{equation}\label{deldef}
n = -m_\beta +\delta\ .
\end{equation}
Notice that in the critical scaling regime, $\delta \asymp L^{-\frac d{d+1}}$ and from the definition of the equimolar radius $r_0$,
\begin{equation}\label{deltaequi}
\delta = 2m_\beta\frac{D_0}{L^d} = 2m_\beta\frac{\sigma_d}d\frac{r_0^d}{ L^d}\ .
\end{equation}

Given any trial function $m(x)$, define $\widehat m(x)$ by truncating $m(x)$ at the levels $h_-$ and $h_+$:
\begin{equation}\label{trunc}
\widehat m(x) = 
\begin{cases} 
h_+  &{\rm  if}\quad  m(x) \ge h_+ \\ 
m(x) 
&{\rm  if}\quad   h_-  < m(x) < h_+  \\
h_- &{\rm  if}\quad  m(x) \le h_- \\ 
\end{cases}\ .\end{equation}

It is clear that
\begin{eqnarray}\label{trunc2}
\int_{A}\int_{A} |\widehat m(x) - \widehat m(y)|^2J(x-y)\dd x\dd y &\le&\int_{A}\int_{A} |m(x) -  m(y)|^2J(x-y)\dd x\dd y \\ &\le& \nonumber 
\int_{{\cal T}_L}\int_{{\cal T}_L} | m(x) - \ m(y)|^2J(x-y)\dd x\dd y \ .\end{eqnarray}

Also clearly, 
$$\int_{{\cal T}_L} F(m(x))\dd x =  \int_A F(m(x))\dd x + \int_B F(m(x))\dd x+\int_C F(m(x))\dd x\ .$$
where $F(m(x))=f(m(x))-f(m_\beta)$. 
Therefore,
\begin{equation}\label{S+B}
\F(m) \ge  \F_S(m) + \F_B(m)
\end{equation}
where
\begin{equation}\label{surp}
\F_S(m) = 
\frac{1}{4} \int_{{\cal T}_L}\int_{{\cal T}_L} |\widehat m(x) - \widehat m(y)|^2J(x-y)\dd x\dd y +
 \int_A F(m(x))\dd x
  \end{equation}
 and
 \begin{equation}\label{bulp}
 \F_B(m) =   \int_B F(m(x))\dd x\ .
 \end{equation}
 
 We shall refer to $\F_S(m)$ as the surface contribution, and to $\F_B(m)$
 as the bulk contribution.  
We shall obtain a lower bound on $f_L(n)$ by separately estimating these contributions.
If the ansatz described in Section 2 is right, then at a minimizing $m$, essentially 
all of the contribution to $\F$ should come from
$\F_S(m)$ and $\F_B(m)$, and so  this lower bound will be quite sharp.

In the next two subsections, we shall estimate $\F_S(m)$ and $\F_B(m)$ separately, starting with $\F_B(m)$. First, however, we close this subsection by showing that if $m$ is any trial function with
$\F(m) < \F(n)$, then $C$ is not empty. (It is clear that if $B$ is empty  and $\kappa>\delta$, as we will assume later on, then the constraint
$\int_{{\cal T}_L}m(x){\rm d}x = nL^d$ cannot be satisfied.)

For this purpose, it is advantageous to rewrite the free energy functional as follows:
Define $\omega$ by $\omega(x) = m(x)-n$. For $m$ satisfying the constraint  (\ref{const}), $\omega$ will satisfy 
\begin{equation}\label{wcon}
\int_{{\cal T}_L} \omega(x){\rm d}x = 0\ .
\end{equation}
Clearly, 
$$\int_{{{\cal T}_L}\times{{\cal T}_L}} |m(x)-m(y)|^2J(x-y){\rm d}x{\rm d}y = \int_{{{\cal T}_L}\times{{\cal T}_L}} |\omega(x)-\omega(y)|^2J(x-y){\rm d}x{\rm d}y\ .$$ Hence, if we  define the functional ${\cal G}$ by
\begin{equation}\label{calGdef}
{\cal G}(\omega) =  \frac{1}{ 4}\int_{{{\cal T}_L}\times{{\cal T}_L}} |\omega(x)-\omega(y)|^2J(x-y){\rm d}x{\rm d}y + \int_{{\cal T}_L} G(\omega){\rm d}x\ ,
\end{equation}
where
\begin{equation}\label{Gdef}
G(\omega) = F(n+\omega)-F(n)- F'(n)\omega
\end{equation}
 we have\begin{equation}\label{FtoG}
{\cal F}(m) = {\cal F}(n) + {\cal G}(\omega)\ ,
\end{equation}
since the term linear in $\omega$ drops out due to (\ref{wcon}) whenever $m$ satisfies the constraint (\ref{const}).

Thus, if  $\F(m) < \F(n)$, then 
${\cal G}(\omega) < 0$, which means that
\begin{equation}\label{Ddef}
D := \{\ x\ :\ G(\omega(x))< 0 \ \} \ne \emptyset\ .
\end{equation}
The next thing to observe is that the set on which $G(\omega)< 0$ is a narrow
interval  $(\omega_-,\omega_+)$ containing $2m_\beta$ whose width is of order $\delta^{1/2}$.
As long as we choose $\kappa$ large compared to $\delta^{1/2}$ (we shall eventually choose $\kappa = \delta^{1/3}$), we will have $D\subset C$.

The reason that this is true can be  seen  in Figure 1, where the function $G(\omega)$ is plotted. 
By  (\ref{Gdef}), one obtains $G$ from $F$ by subtracting the tangent line to the graph of $F$ at $m= n$ away from $F$, and then changing variables from $m$ to $\omega$, which measures deviations from $n$.

\begin{figure}
\begin{center}
\includegraphics[width=8.5cm,height=4.5cm]{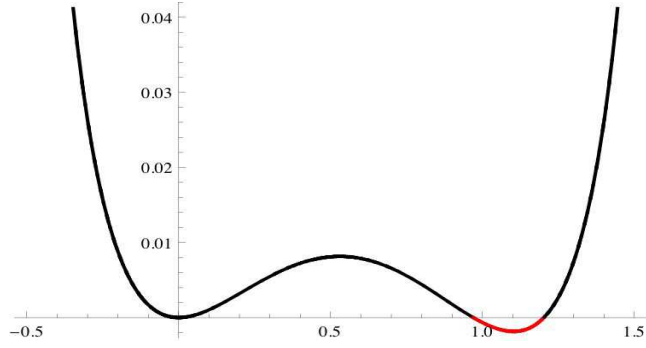} 
\caption{Plot of the function $G(\omega)$}
\end{center}
\end{figure}

Since $F$ is locally convex near $m = n$, $G$ is locally convex near $\omega = 0$, and so 
$\omega =0$ is one local minimum of $G$. Subtracting off the tangent line function ``tilts'' the graph of $F$ downward to the right, and so the minimum of $F$ at $m = m_\beta$ gets tilted to  a slightly negative value; the global minimum of $G$ lies in this dip below the axis. Since $F'(n) = {\cal O}(\delta)$,
the (negative) value of the global minimum is on the order of $-2m_\beta  F'(n) $; i.e., also  
${\cal O}(\delta)$. But since the curvature at the minimum is strictly positive, the width of the interval on which $G$ is negative is ${\cal O}(\delta^{1/2})$.

The following lemma makes this, and more,  precise:

\medskip
\begin{lm} \label{lb2}  Let $G$ be defined by (\ref{Gdef}). Then the equation $G(\omega) = 0$ 
has exactly three solutions, $0, \omega_-$ and $\omega _ +$ where
$0 < \omega_- < \omega _ +$. There is a constant $c$ such that for all $L$ sufficiently large
\begin{equation}\label{top21}
|\omega_\pm - 2m_\beta| \le c\delta^{1/2}\ ,
\end{equation}
and, $G(\omega) < 0$ if and only if $\omega\in (\omega_-,\omega_+)$.
Consequently, if $m$ is any trial function with $\F(m) < \F(n)$, and if $\kappa = \delta^{1/3}$ is used in 
the definition of (\ref{hpmdef}), and hence of $C$, then  $C$  is not empty.

Moreover, $G$ has a unique global minimizer $\omega_\star$ which satisfies
$$|n + \omega_\star - m_\beta| \le c\delta$$
for some fixed constant $c$ depending only on $F$. 
If $m$ is any trial function with $\F(m) < \F(n)$, there is an $a < n$ so that
$$m_{{\rm trunc}}(x) := \max\{\ a\ , \ \min\{\ n+\omega_\star, m(x)\ \}\ \}$$
is also a valid trial function with $\F(m_{{\rm trunc}}) \le \F(m)$, and the sets $A$, $B$ and $C$ determined by $m_{{\rm trunc}}$ are the same as those determined by $m$. 
\end{lm}
\medskip

\begin{remark}\label{sharper}
 The last part of this lemma says that if we seek to prove any theorem concerning the sets $A$, $B$ and $C$ associated to a trial function $m$ with $\F(m) < \F(n)$, then we may freely assume that $m$ is bounded above by  $n+\omega_\star = m_\beta+{\cal O}(\delta)$.  It also implies that any {\em minimizer} $m$ must be bounded above by $m_\beta+{\cal O}(\delta)$. 
\if false 
However, a sharper statement for minimizers can be proven using the Eluler-Lagrange equation; see the final section. \fi
\end{remark}

\noindent{\bf Proof of Lemma~\ref{lb2}:}  For the first part, we simply provide formulas that quantify the remarks in the paragraph preceding the statement of the lemma.  It is clear from (\ref{Gdef}) that $\omega = 0$ is one solution of $G(\omega) = 0$. Also, since $F'(n) = {\cal O}(\delta)$, and 
$F(n) = {\cal O}(\delta^2)$, it follows that for some $c<\infty$,
$$G(\omega) \ge F(n+\omega) - c\delta\ .$$
Thus, $G(\omega)< 0$ requires $F(n+\omega) < c\delta$; i.e., $n+\omega$ must lie in one of the
two ``wells'' of $F$.  As before, let $1/\chi$ denote $F''(\pm m_\beta)$, the second derivative of $F$
at the bottom of the two wells. Let $\ell$ be defined by
$$\ell = \inf\left\{\ m>0\ :\ F''(m) \ge \frac{1}{2\chi}\ \right\}\ .$$
Then $F(m) \ge F(\ell) > 0$ on $[-\ell, \ell]$, and so 
$$G(\omega) \ge F(\ell) - c\delta\ $$
for $|n+\omega| < \ell$.  For $L$ large enough, $F(\ell) - c\delta>0$, and so we may restrict our attention to values of $\omega$ in the intervals $-1\le n+\omega \le -\ell$ and $\ell \le n+\omega \le 1$. 
$G$ is strictly convex in both of these intervals, and since $G'(n) = 0$, $\omega = 0$
is the unique minimizer of $G$ in the left interval.

To find the solutions in the right interval, 
introduce a new variable $u$ 
defined $u$ by $n+\omega=: -n-u$ and define $H(u): =G(-2n-u)$. Then $\ell \le n+\omega \le 1$
if and only if $-(n+1) \le u \le -(n+\ell)$, and for such $u$
 \begin{eqnarray}\label{roots}
 H(u)&=& F(-n-u)-F(n) +2F'(n)n +F'(n) u\nonumber\\
 &=&\frac{1}{2}F''(-n-\xi)u^2+2F'(n) u +2F'(n)n\nonumber\\
 &\ge&  \frac{1}{4\chi}u^2 + 2F'(n)n +F'(n) u\nonumber\\
 &=&  \frac{1}{4\chi}(u+ 4\chi F'(n))^2 + 2F'(n)n -4\chi(F'(n))^2\ ,
 \end{eqnarray}
 with $\xi\in[\min\{0,u\},\max\{0,u\}]$ in  the second line. We have used the Taylor expansion  $F(-n-u)=F(-n)-F'(-n)u+\frac{1}{2}F''(-n-\xi)u^2$ and the fact that $F(z)$ is even and $F'(z)$ is odd, and finally, the lower bound on $F''$
 in the well. 
 
Evidently,
$H(u)> 0$
unless
$u_-  < u <  u_+$
where $u_\pm$ are the two roots of the quadratic expression on the right in (\ref{roots}).
Since $F'(n) = {\cal O}(\delta)$, it is evident that there is a constant $c$ such that
$$|u_\pm| \le c\sqrt\delta\ .$$

By the local convexity of $G$, the remaining two solutions of $G(\omega) =0$ must lie in the corresponding interval; i.e., $(-2n -u_+, -2n-u_-)$, and $G$ is positive outside this interval. except at
$\omega = 0$ 

When $\F(m) < \F(n)$,  $m$ is not constant, and  $\displaystyle{\int_{{{\cal T}_L}\times{{\cal T}_L}} |\omega(x)-\omega(y)|^2J(x-y){\rm d}x{\rm d}y> 0}$. Hence, for $\F(m) < \F(n)$, we must have $G(\omega(x))< 0$ on a set of positive measure.  This proves the statements made in the first paragraph of the lemma. 

To prove the claims made in  the second paragraph, note that $G(\omega)$ is an increasing function of $\omega$
to the right of its global minimum  $\omega_\star$.
Since truncation always lowers the interaction energy
$\displaystyle{\int_{{{\cal T}_L}\times{{\cal T}_L}} |\omega(x)-\omega(y)|^2J(x-y){\rm d}x{\rm d}y}$,
replacing $\omega(x)$ by $\min\{\ \omega(x)\ , \ \omega_\star\ \}$ lowers ${\cal G}(\omega)$.
Note that, if $\omega(x)$ was not already bounded above by $\omega_\star$, the truncated function
will no longer satisfy the constraint (\ref{wcon}). 

However, we can remedy this by another truncation at the other end: Let $[\omega]_\pm$
denote the positive and negative parts of $\omega$. Then by  (\ref{wcon}),
$$\int_{{\cal T}_L}[\omega]_- {\rm d}x = \int_{{\cal T}_L}[\omega]_+ {\rm d}x$$
and since 
$$a \mapsto  \int_{{\cal T}_L}\min\{\ [\omega]_-\ , \ a\ \} {\rm d}x$$
increases continuously from $0$ to $\int_{{\cal T}_L}[\omega]_- {\rm d}x$ 
as $a$ increases from $0$ to $1$, we can choose $a$ so that
$$\int_{{\cal T}_L}\min\{\ [\omega]_-\ , \ a\ \} {\rm d}x = 
\int_{{\cal T}_L}\min\{\ [\omega]_+\ , \ \omega_\star\ \} {\rm d}x\ .$$
Since $G(\omega)$ is a decreasing function of $\omega$ on $(-\infty,0)$, this second truncation also
lowers ${\cal G}(\omega)$, and restores the constraint (\ref{wcon}). Finally, truncating
$\omega$ at $-a$ and $\omega_\star$, corresponds to a truncation in $m$ as in the lemma. 
The bound on $|n+\omega_\star - m_\beta|$ comes from the fact that $\omega_\star$ is the unique
non-zero 
solution to 
$$F'(n+\omega) = F'(n)\ ,$$
and what we have said in the proof of the first paragraph. 
\qed

\begin{remark}\label{trivial} 
As mentioned in the introduction, our minimization problem is trivial for
$n\in [-1,-m_\beta]$ (and hence for $n\in [m_\beta,1]$ by symmetry). This is easily seen by considering the functional ${\cal G}$ used in the previous proof: If  $n\in (-1,-m_\beta)$, then the tangent line being subtracted from the graph of $F$ in (\ref{Gdef}) would have a {\em negative slope}, and so subtracting
it off would tilt the graph upward to the right, and not downward. Hence $G$ will have a unique
global minimum at $\omega =0$. 
Thus the unique minimizer of ${\cal G}$ is the constant profile $\omega = 0$, and
then by (\ref{FtoG}) the unique minimizer of ${\cal F}$ is the constant profile $m = n$.  For the values $n =-1$ and $n=-m_\beta$, the situation is even more elementary. 
\end{remark}

\subsection{The bulk contribution} %\label{lb5}

The key to estimating  $\mathcal F_B$ is that for $\kappa$ small enough,  $F$ is strictly convex on 
 $(-1,h_-)$:
For any $h\in (-1,h_-)$, $\displaystyle{F''(h)\ge -1+\frac{1}{\beta}\frac{2}{1-h_-^2}}$. Define the quantity $\chi_-$ by
$$\frac{1}{ \chi_-} = F''(h_-) = -1+\frac{1}{\beta}\frac{2}{1-h_-^2}\ .$$
Then, by Taylor's Theorem, and using the fact that $F(-m_\beta) = F'(-m_\beta) = 0$, we have
$$F(m(x)) \ge \frac{1}{ 2 \chi_-}(m(x) + m_\beta)^2$$
everywhere on $\{ x\in {\cal T}_L\,|\,m(x) \le h_-\}$.

Therefore,
\begin{eqnarray}\label{tou}
\int_B F(m(x)){\rm d}x &=&   |B|\left(\frac{1}{ |B|}
\int_{B} F(m(x)){\rm d}x\right)\nonumber\\
&\ge&
 |B|\frac{1}{ 2\chi_-}\left(\frac{1}{ |B|}
\int_{B} (m(x)+m_\beta)^2{\rm d}x\right)\\
&\ge& 
 |B|\frac{1}{ 2\chi_-}\left(\frac{1}{ |B|}
\int_{B} (m(x)+m_\beta){\rm d}x\right)^2\nonumber\\
&=& \frac{1}{ 2\chi_- |B|}\left(
\int_{B} m(x){\rm d}x + m_\beta |B|\right)^2\ .\nonumber
\end{eqnarray}

This estimate should be quite sharp, since we expect any nearly minimizing profile $m(x)$ to be nearly constant
in $B$. Our problem is now reduced  to that of estimating  $\int_{B} m(x){\rm d}x$. Before going into the details, let us summarize what we would expect, and what lemmas we shall need, to prove what we would expect.

First, we would expect the transition region $A$ to be very ``thin'', so that:

\smallskip
\noindent{$\bullet$}{\em  $|A|$ is negligible compared to
$|B|$ and $|C|$.}
\smallskip

 In that case, we would expect
\begin{eqnarray}
\int_{B} m(x){\rm d}x &\approx&  \int_{{\cal T}_L} m(x){\rm d}x -  \int_{C} m(x){\rm d}x\nonumber\\
&=& L^d(-m_\beta+\delta) -  \int_{C} m(x){\rm d}x\ , \nonumber\\
&\approx& L^d(-m_\beta+\delta) -  \frac{\sigma_d}{d} R^dm_\beta\ ,\label{tou2} \end{eqnarray}
where in the last line, we have used the fact that $m(x)$ is very close to $m_\beta$ on $C$. In fact, 
on $C$, $m(x) \ge m_\beta-\kappa$ by definition, and Lemma~\ref{lb2} and the remark following it will allow us to assume an upper bound of the form $m_\beta+{\cal O}(\delta)$. 
Also, if $|A|$ is negligibly small,
\begin{equation}\label{tou3}
|B|m_\beta \approx  (L^d - \frac{\sigma_d}{d} R^d)m_\beta\ .
\end{equation}

Using  (\ref{tou3}) and (\ref{tou2}), we would have
${\displaystyle \int_{B} m(x){\rm d}x + m_\beta |B| \approx  L^d(\delta  - 2m_\beta \frac{\sigma_d}{d}  (R/L)^d)}$.
Then using  (\ref{tou})  together with the simple (but not extravagant) bound $|B| < L^d$, we would have
\begin{equation}\label{tou5}
{\cal F}_B(m) \gtrapprox \frac{1}{2\chi_-L^d}(\delta L^d - 2m_\beta\frac{\sigma_d}{d} R^d)^2\ .
\end{equation}
The next lemma gives the precise statement:
\begin{lm}\label{lb5}
 Let $m$ be any trial function such that ${\cal F}(m)\le {\cal F}(n)$, and such that $m$ is bounded above by $n+\omega_\star$, where $\omega_\star$ is defined and estimated in Lemma~\ref{lb2} . Then
$$\mathcal F_B(m)\ge  \frac{L^d}{ 2\chi_- }\left(\left(\delta  - 2m_\beta\frac{\sigma_d}{d}  \frac{R^d}{L^d}\right)^2 - \e\right) \ ,$$
with $\e$ given by 
\begin{equation}\label{tou6}
\e =  4\Big| \delta     - 2m_\beta\frac{\sigma_d}{d}  \frac{R^d}{L^d}\Big |\left(2m_\beta c\frac{\delta^2}{\kappa^2} + 
\frac{\sigma_d}{d}  \frac{R^d}{L^d}(\kappa+c\delta)\right)
\end{equation}
for some  constant $c$.
\end{lm}

\begin{remark} Note that if we choose $\kappa = \delta^{1/3}$, and if $R^d/L^d = {\cal O}(\delta)$, then
$\epsilon = {\cal O}(\delta^{7/3})$. The surface contribution will limit the side of $R$, preventing cancelation in the main term, so that it will be ${\cal O}(\delta^{2})$, and hence strictly larger. 
\end{remark}

To prove Lemma~\ref{lb5} 
we first need to show that $|A|$ is in fact negligible, as explained in the heuristics. The following Lemma
takes care of that:

\medskip
\begin{lm}\label{lb4}
Let $m$ be any trial function such that ${\cal F}(m)\le {\cal F}(n)$. Then, for some finite $c>0$ and $c'>0$ 
$$|A| \le \frac{ F(n)}{ c'\kappa^2} \le  c \frac{\delta^2}{ \kappa^2}L^d\ .$$
\end{lm}
\medskip

\noindent{\bf Proof:}   We have 
$$ F(h_+)=c' \kappa^2=F(h_-)$$
where $c'=\frac 1 2F''(p)$, for some $p$ with $m_\beta-\kappa\le p \le m_\beta$ and $c'>0$ for $\kappa>0$ small enough. 
  It is easy to see from  the definition of $A$ and the properties of the function $F$
that  uniformly on $A$, 
$$F(m(x)) \ge F(h_+)=c' \kappa^2\ .$$
Therefore
$$I_A \ge |A|c' \kappa^2 \ .$$
On the other hand, since ${\cal F}(m)\le {\cal F}(n)$, 
$$I_A < {\cal F}(n) = F(n)L^d\ .$$
Since $F(n)=c_1 \delta^2$, 
we get the result. \qed

 \medskip
 
Now that we have Lemma~\ref{lb4}, we return to the proof of Lemma~\ref{lb5}:

\medskip

\noindent{\bf Proof of Lemma~\ref{lb5}:} Note that
$$
\int_{B} m(x){\rm d}x = nL^d  - \int_{C} m(x){\rm d}x - \int_{A} m(x){\rm d}x\ .$$
By Lemma \ref{lb4}, 
$$-m_\beta |A| \le h_-|A| \le   \int_{A} m(x){\rm d}x  \le h_+|A| \le m_\beta|A|\ .$$
Thus,

$$\left| \int_{B} m(x){\rm d}x - (nL^d - |C|m_\beta )\right| \le m_\beta |A| + \left| |C|m_\beta -\int_{C} m(x){\rm d}x\right| \ .$$
Since $n=-m_\beta +\delta$ and it is evident that $|B| = L^d -\displaystyle{\frac{\sigma_d}{d}}  R^d  - |A|$, 
\begin{equation}\label{blue1}
\left| \left(\int_{B} m(x){\rm d}x + m_\beta |B|\right) - (\delta   L^d  - 2m_\beta|C|) \right| \le 2m_\beta|A| + 
\left| |C|m_\beta -\int_{C} m(x){\rm d}x\right|\ .
\end{equation} 

Next,  on $C$, $m(x) \ge m_\beta -\kappa$ by the definition of $C$. Also, by the hypothesis that 
$m(x) \le n+\omega_\star$ for all $x$, and Lemma~\ref{lb2},  $m(x) \le m_\beta + c\delta$ for some fixed constant $c$, and for all $x$. Thus,
$$\left| |C|m_\beta -\int_{C} m(x){\rm d}x\right| \le |C|(\kappa + c\delta)\ ,$$
and hence, (\ref{blue1})  yields
$$
\left| \left(\int_{B} m(x){\rm d}x + m_\beta |B|\right) - (\delta   L^d  - 2m_\beta|C|) \right| \le 2m_\beta|A| + 
 |C|(\kappa + c\delta)\ .
$$

Since the inequality $|a-b|\le c$ implies $a^2\ge b^2-2|b|c$, 
we have
\begin{multline}\left(\int_{B} m(x){\rm d}x + m_\beta|B|\right)^2 \ge\\ \Big(\delta L^d - 2m_\beta|C|\Big)^2 - 2\Big| \delta   L^d  - 2m_\beta|C|\Big|\left(2m_\beta|A| + 
 |C|(\kappa + c\delta)\right) \ .\end{multline}
Going back to (\ref{tou}) and using the estimate $|B| < L^d$ and the definition $|C| = \sigma_dR^d/d$, we obtain
\begin{eqnarray}\label{tou99}
{\cal F}_B(m)&\ge& \frac {L^d}{2\chi_-}\left(\delta  - 2m_\beta\frac{\sigma_d}{d}  \frac {R^d} {L^d}\right)^2  \nonumber \\ &- &\frac {L^d} {2\chi_-} 4\Big| \delta     - 2m_\beta\frac{\sigma_d}{d}  \frac{R^d}{L^d}\Big |\left(2m_\beta \frac{|A|}{L^d} + 
\frac{\sigma_d}{d}  \frac{R^d}{L^d}(\kappa+c\delta)\right)\ .
\end{eqnarray}
Now using Lemma~\ref{lb4} to estimate $|A|$, we obtain the result. 
\qed

\subsection{The surface  contribution} %\label{lb4}

Our goal in this subsection is to prove the following estimate.

\medskip
\begin{lm}\label{lbA1}  Let $m$ be any trial function such that ${\cal F}(m)\le {\cal F}(n)$, 
\begin{equation}\label{tou20}
\F_S(m) \ge \left[1- {\cal O}\left(\frac{1}{R-2}\right)\right]_+\sigma_d[R-2]_+^{d-1}\left(1-\frac{\kappa}{m_\beta}\right)^2S 
\end{equation}
where $S$ is the surface tension, and $[a]_+ = \max\{\ a\ , \ 0\ \}$, so that the bound is trivially true for $R<2$, twice the range of $J$. 
\end{lm}
\medskip

Let us first explain the heuristics, and then collect the lemmas required to substantiate them. 
To prove this lemma we need to relate ${\cal F}_S$ to the one dimensional functional defined in (\ref{pst}), which gives the planar surface tension $S$. To do this, we use rearrangement inequalities to replace our near minimizer $m$ by a radial function on all of $\mathbb{R}^2$. This radial function will give us a trial function for (\ref{pst}). 

The Riesz rearrangement inequality that we intend to use
applies to functions on $\R^d$, and not on the torus, hence the first thing we have to do is to extend ${\cal F}_S$ to a functional on profiles in all of $\R^d$ without lowering the value of ${\cal F}_S$ too much. 
Here is why we can expect that this is possible.

We expect that for a non-constant near minimizer $m$, 
$C$ should  be essentially a sphere of radius $R$ that we can take to be centered in ${\cal T}_L$, considered as a  $d$-cube of side length $L$ in $\R^d$, and that the whole transition region $A$ will be in an annulus close to $C$. In particular, the truncation $\widehat m$ of $m$ that is defiend in (\ref{trunc})
and used in the definition (\ref{surp}) of $\F_S$ satisfies
$\widehat m(x) = h_-$
for all $x$ within unit distance of  the boundary of the square.  Now 
extend $\widehat m$
to a function $\widetilde m$ on all of $\R^d$. Do this by defining $\widetilde m(x) = h_-$ for
$x$ outside ${{\cal T}_L}$; i.e.,
 \begin{equation}\label{ext}
 \widetilde m(x) = 
\begin{cases} 
\widehat m(x)  &{\rm  if}\quad  x \in {\cal T}_L \\ 
h_- &{\rm  if}\quad  x\in \R^d\backslash {\cal T}_L\ . \\ 
\end{cases}\ .
 \end{equation}

Then, since $J$ is supported by the unit sphere, it would follow from $\widetilde m(x) = h_-$
everywhere near the boundary of ${\cal T}_L$ that 
\begin{equation}\label{tou8} 
\int_{{\cal T}_L}\int_{{\cal T}_L} |\widehat m(x) - \widehat m(y)|^2J(x-y)\dd x\dd y =
  \int_{\R^d}\int_{\R^d} |\widetilde m(x) - \widetilde m(y)|^2J(x-y)\dd x\dd y\ .
\end{equation}
  As for the potential term, define  
$\widetilde F$ by
\begin{equation}\label{tou9} 
\widetilde F(m) = 
\begin{cases} 
F(m)  &{\rm  if}\quad  h_+> m > h_- \\ 
0  &{\rm  if}\quad  m \le h_- \ \text{ or} \ m\ge h_+\\ 
\end{cases}\ .
\end{equation}
With this definition,
\begin{equation}\label{tou10} 
\int_{A}F(m(x))\dd x =  \int_{A}F(\widehat m(x))\dd x=
 \int_{\R^d}\widetilde F(\widetilde m(x))\dd x\ . 
 \end{equation}
Then combing (\ref{surp}), (\ref{tou8}) and (\ref{tou10}), we would have
$$\F_S(m) \ge   
 \int_{\R^d}\widetilde F(\widetilde m(x))\dd x + 
\frac{1}{4} \int_{\R^d}\int_{\R^d} |\widetilde m(x) - \widetilde m(y)|^2J(x-y)\dd x\dd y\ .$$
 
We are now in a position to use rearrangement inequalities to make contact with the one dimensional variational problem (\ref{pst}) that defines the planar surface tension:
Let $m^*$ denote the spherical decreasing rearrangement of $\widetilde m$ (see \cite{LL}). Then by the Riesz rearrangement inequality \cite{LL},
$$  \int_{\R^d}\int_{\R^d} |\widetilde m(x) - \widetilde m(y)|^2J(x-y)\dd x\dd y \ge
 \int_{\R^d}\int_{\R^d} | m^*(x) -  m^*(y)|^2J(x-y)\dd x\dd y\ ,$$
 and of course
 $$ \int_{\R^d}\widetilde F(\widetilde m(x))\dd x  =  \int_{\R^d}\widetilde F(m^*(x))\dd x\ .$$
 Therefore, if our intuition about the size and shape of $C$ is right, we should have
 $$\F_S(m) \ge 
 \frac{1}{4}\int_{\R^d}\int_{\R^d} | m^*(x) -  m^*(y)|^2J(x-y)\dd x\dd y + 
 \int_{\R^d}\widetilde F(m^*(x))\dd x\ .$$
 Let $r = |x|$.   Because of the spherical rearrangement, $m^*(x)$ depends on $|x|$, and the  corresponding region $C$ is indeed a sphere of radius $R$. 
 
 The key to making contact with the planar surface tension is the fact that 
 $m^*(r)/(1-\frac{\kappa}{m_\beta})$, extended by $m_\beta$ for $r<0$, is a valid trial function
 for the one dimensional variational problem (\ref{pst}) defining $S$. We shall use this fact to get a lower bound on $\F_S(m)$ for a non-constant minimizer $m$ that is of the form (\ref{tou20}). Note that
 apart from some small corrections, the main term in this bound is $\sigma_d R^{d-1}S$, the contribution we would expect for a droplet of radius $R$. 
 
 To carry out this program of estimation, we need to  show that $|C|$ is not too large: If $|C|$
is large, then it is easy for $C$ to ``wrap around'' so that (\ref{tou8}) is not even approximately true. 
Then we would be prevented from applying the Riesz rearrangement inequality. Then next  lemma
shows that in fact, if $m$ is a trial function with $\F(m) < \F(n)$, then $C$, which we know to be non-empty by Lemma~\ref{lb2},  has volume $|C| \le {\cal O}(L^d\delta)$. 

\medskip
\begin{lm}\label{lb4B}
Let $m$ be any trial function such that ${\cal F}(m)\le {\cal F}(n)$.  Then, for $L$ sufficiently large,  there is a constant $c$ such that 
$$|C| \le cL^d\delta=cL^{\frac{d^2}{d+1}}\ .$$
\end{lm}
\medskip

\noindent{\bf Proof:}   
Since 
\begin{equation}\label{obv}
nL^d= (-m_\beta+\delta)L^d=\int_A m(x) \dd x+\int_B m(x) \dd x+\int_C m(x) \dd x,\end{equation}
we can use the obvious lower bounds $m(x)\ge -1$ on $A\cup B$ and $m(x)\ge h_+$ on $C$ to conclude that
$$(-m_\beta+\delta)L^d\ge -(|A|+|B|)+ (m_\beta-\kappa)|C|.$$
By using $|C|=L^d-(|A|+|B|)$, we get
$$(|A|+|B|)(1+m_\beta -\kappa)\ge (2m_\beta-\delta)L^d.$$
By Lemma \ref{lb4}, $|A|\le c\delta^2\kappa^{-2}L^d$. Hence, for $L$ sufficiently large
$$|B|\ge \frac{m_\beta}{1+m_\beta}L^d.$$
On the other hand, by (\ref{tou})
$${\cal F}(n)\ge{\cal F}(m)\ge\int_B F(m)\dd x\ge \frac{|B|}{2\chi_-}\left(\frac{1}{|B|}\int_B m(x)\dd x +m_\beta\right)^2.$$
Therefore
$$\left|\frac{1}{|B|}\int_B m(x)\dd x +m_\beta\right|\le \sqrt{\frac{2\chi_-{\cal F}(n)}{|B|}}\le\delta\sqrt{2\chi_-c_1\frac{1+m_\beta}{m_\beta}}=c_2\delta,$$
by using ${\cal F}(n)\le c_1\delta^2L^d$.

Finally, using this in (\ref{obv}) we have:
$$(-m_\beta+\delta)L^d\ge -(L^d-|A|-|C|)\frac{1}{|B|}\int_B m(x)\dd x-|A|+ (m_\beta-\kappa)|C|.$$
Then
$$|C|(2m_\beta-\kappa-c_2\delta)\le L^d(\delta+c_2\delta)+|A|.$$
Using Lemma \ref{lb4} to bound $|A|$,  and taking $L$ sufficiently large we conclude the proof.\qed

Armed with this lemma on $|C|$, we return to the proof of Lemma~\ref{lbA1}. In our heuristic discussion, we relied on our expectation that $C$ is nearly a disk centered in ${\cal T}_L$ 
 (with an appropriate choice of the origin in ${\cal T}_L$) with a radius small compared to $L$
 in order to justify (\ref{tou8}). At this stage something less can be proved, which still suffices for the proof of Theorem \ref{thm1}:
 
  \smallskip
 \noindent{$\bullet$}{\em If $m$ is a trial function with $\F(m) < \F(n)$,  and  if 
 $C$ does  ``wrap around the torus'' ${\cal T}_L$, then the arms that
 ``wrap around'' are very thin, as shown in Figure 2.}

\begin{figure}
\centering \includegraphics[width=8cm,height=8cm]{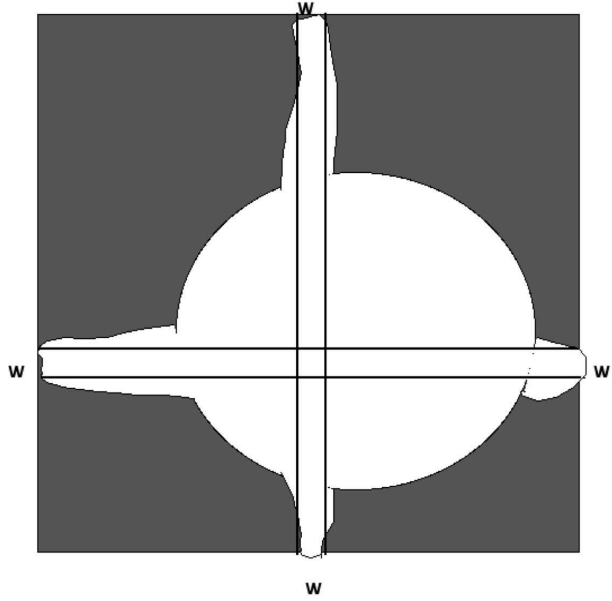} 
\caption{Possible droplet shape.}
\end{figure}

 For any of the $d$ coordinate directions $1 \le i \le d$,  consider the volume of $A\cup C$ that is contained in the slab $a \le x_i \le a+2$.
 If for each choice of $a$ this volume is at least $w$, then 
 $$|A\cup C| \ge \frac{L}{2}w\ ,$$
 since integrating this volume in $a$ from $-L/2$ to $L/2$ gives twice the volume of  $|A\cup C|$. Hence there is at least one choice of $a$ for which 
 $$|\{\  x\ :\ a\le x_i \le a+2\ \}\cap (A\cup C)|  \le \frac{2|A\cup C|}{L}\ .$$
Now by the translation invariance of $\F$, we may freely translate $m$, and so may assume that $a = L/2$. 
Thus without loss of generality, we may assume that for each coordinate direction $i$, 
\begin{equation}\label{ends}
\big| \left \{ \ x\ :\ |x_i \pm L/2| \le 1\ \right\}\cap (A\cup C)\big|  \le \frac{2|A\cup C|}{L}\ .
\end{equation}

\medskip
\begin{lm}\label{lbA} 
There is a constant $c$ so that for any trial function $m$ with $\F(m) < \F(n)$, 
$$ \int_{{\cal T}_L}\int_{{\cal T}_L} |\widehat m(x) - \widehat m(y)|^2J(x-y)\dd x\dd y \ge
  \int_{\R^d}\int_{\R^d} |\widetilde m(x) - \widetilde m(y)|^2J(x-y)\dd x\dd y -   2d\frac{|A\cup C|}{L}
 \ ,$$
 and for some finite $c$, 
 $$
2d  \frac{|A\cup C|}{L}  \le cL^{d-1}\left(\delta+ \frac{\delta^2}{\kappa^2}\right)\ .
$$
\end{lm}
\medskip

\noindent{\bf Proof:} The integral on the left hand side can only be smaller than the integral on the right hand side only 
on account of pairs of points $(x,y)$ with, say, $x$ in the cube representing ${\cal T}_L$, and $y$ outside it, and where $x\in A\cup C$, since otherwise $m(x) = m(y) = h_-$.  Since $J$ has unit range, $x$ must have
$|x_i \pm L/2|\le 1$ for at least one $1 \le i \le d$. Hence the total contribution from such pairs of points
is a fixed multiple of $|A\cup C|/L$, by  (\ref{ends}). Then using our bounds on $|A|$ and $C$ from Lemmas~\ref{lb4}
and \ref{lb4B}, we obtain final bound. 
\qed

\medskip

Lemma~\ref{lbA} gives us the rigourous replacement for (\ref{tou8}) in our heuristic analysis. We now
turn to the term involving $\widetilde{F}$. Note that   $|x|\ge R$ at all points $x$ in the support of $\widetilde F(m^*(x))$. Hence, going to spherical  coordinates, 
 \begin{equation}\label{potter}
 \int_{\R^d}\widetilde F(m^*(x))\dd x = 
 \sigma_d \int_{R}^\infty\widetilde F(m^*(r))r^{d-1}\dd r \ge 
\sigma_d R^{d-1}\int_{R}^\infty\widetilde F(m^*(r))\dd r\ .
 \end{equation}
 
To proceed, we once again use (\ref{curve}). By construction, $m^*(r) = m_\beta - \kappa$ for all
$$r \le \left(\frac{d}{\sigma_d}(|A|+|C|)\right)^{\frac 1 d}$$
and the r.h.s. is larger than $R$.

Therefore, if $r < R-1$, then 
\begin{equation}\label{zero} | m^*(r) -  m^*(s)|^2\overline J (r-s) = 0\end{equation}
for all $s$, since $\overline J$ has unit range. 

Hence by  (\ref{curve}),
\begin{multline}\label{tou300}
\int_{\R^d}  \int_{\R^d}  | m^*(x) -  m^*(y)|^2 J (x-y) \dd x\dd y =\\
\left[1- {\cal O}\left(\frac{1}{R-2}\right)\right]_+\int_0 ^\infty  \int_0 ^\infty  | m^*(r) -  m^*(s)|^2\overline J (r-s)\sigma_d r^{d-1}\dd r\dd s\ ,\end{multline}
where once again, $[\cdot]_+$ is the positive part function. Indeed, the left hand side is clearly positive, so if $R$ is sufficiently small that using the uniform bound on the Jacobian that led to (\ref{curve}) provides a negative lower bound, simply use the trivial lower bound by zero instead.

Next, since (\ref{zero}) implies $\overline J (r-s)r^{d-1}\ge [R-1]_+^{d-1}\overline J (r-s)$ for all $r,s\ge 0$,  
we have
%$$ | m^*(r) -  m^*(s)|^2\overline J (r-s) = 0 \qquad{\rm for}\qquad r < R-1\ ,$$
%and so 
\begin{multline}\label{tou333}
\int_{\R^d}  \int_{\R^d}  | m^*(x) -  m^*(y)|^2 J (x-y) \dd x\dd y =\\
\left[1- {\cal O}\left(\frac{1}{R-2}\right)\right]_+[R-1]_+^{d-1}\int_0 ^\infty  \int_0 ^\infty  | m^*(r) -  m^*(s)|^2\overline J (r-s)\sigma_d \dd r\dd s\ ,\end{multline}

\medskip
We are now
ready to prove Lemma~\ref{lbA1}:

\medskip

\noindent{\bf Proof of Lemma~\ref{lbA1}:}  By (\ref{tou10}) and (\ref{potter})
 for the terms involving $F$, and Lemma~\ref{lbA} and (\ref{tou333}), it remains only to show that
$$\frac{1}{4}\int_0 ^\infty  \int_0 ^\infty   | m^*(r) -  m^*(s)|^2\overline J (r-s)\dd r\dd s + 
 \int_0^\infty \widetilde F(m^*(r))\dd r \ge (1- \frac{\kappa}{m_\beta})^2S\ .$$
 Our proof of this rests on the fact that $m^*(r)/(1-\frac{\kappa}{m_\beta})$, extended by $m_\beta$ for $r<0$, is a valid trial function
 for the variational problem defining $S$.  To show this, we need to prove that it approaches the correct asymptotic values; i.e., that it satisfies the constraint imposed on (\ref{pst}).
 In fact, due to the  lemma \ref{lb2},  $C$ in not empty when $\F(m) < \F(n)$, and of course $B$ is not empty too. So, by the definition of $C$, $m^*(r) = m_\beta - \kappa$ for all $r < R$.
Consequently,  $m^*(r)-m^*(s)=0$ if $ r,s<R$ so that {\em provided $R>1$, the range of $J$} we can extend the region of integration to $(-\infty,+\infty)$: 
\begin{multline}\int_0 ^\infty  \int_0 ^\infty   | m^*(r) -  m^*(s)|^2\overline J (r-s)\dd r\dd s= \\
 (1-\frac{\kappa}{m_\beta})^2\int_{\R}  \int_{\R}
  | m^*(r)/(1-\frac{\kappa}{m_\beta}) -  m^*(s)/(1-\frac{\kappa}{m_\beta})|^2\overline J (r-s)\dd r\dd s\ .\end{multline}
Now, if $R\le2$, the bound in the Lemma is trivial, and there is nothing to prove, Hence there is no harm in assuming that $R>2$, which we now do. 
   
 Since $\widetilde F$ is decreasing on $m>0$, and since for $m>0$, $m/(1-\frac{\kappa}{m_\beta}) > m$,
 it follows that
 $$\widetilde F(m^*(x))\ge F\left(m^*(x)/(1-\frac{\kappa}{m_\beta})\right) \ge (1-\frac{\kappa}{m_\beta})^2 F\left(m^*(x)/(1-\frac{\kappa}{m_\beta})\right)\ .$$
 Therefore,
 \begin{eqnarray}
&{\phantom 1}&\frac{1}{4}\int_0 ^\infty  \int_0 ^\infty   | m^*(r) -  m^*(s)|^2\overline J (r-s)\dd r\dd s +
 \int_0^\infty \widetilde F(m^*(r))\dd r \ge \nonumber\\
 &{\phantom 1}&\frac{1}{4}(1-\frac{\kappa}{m_\beta})^2\left[
 \int_{\R}  \int_{\R}
  | m^*(r)/(1-\frac{\kappa}{m_\beta}) -  m^*(s)/(1-\frac{\kappa}{m_\beta})|^2\overline J (r-s)\dd r\dd s\right. \\ &+&\left.\int_\R F\Big(m^*(r)/(1-\frac{\kappa}{m_\beta})\Big)\dd r\right] \nonumber
\\&\ge&(1-\frac{\kappa}{m_\beta})^2 S\ .\nonumber
  \end{eqnarray}
  The proof of Lemma \ref{lbA1} is complete.  \qed
 
\medskip

\medskip

\medskip
\section{Proofs of the theorems.}

\subsection{Proof of the Theorem \ref{thm1}} We fix the value $\kappa = \delta^{1/3}$ for the proof. Given a trial function $m$ with $\F(m) < \F(n)$, we replace $m$ by its truncation as defined in Lemma~\ref{lb2}. This
 lowers the free energy, and does not change the sets 
$A$, $B$ and $C$.  In summary, after this replacement we have a trial function that has a free energy at least as low as the one we started with, the same value of $R$, and which is bounded above by $n +\omega_\star$ as in Lemma~\ref{lb2}. 

Then using 
 Lemmas \ref{lbA1} and \ref{lb5} we conclude that
\begin{eqnarray}
{\cal F}(m) &\ge& {\cal F}_S(m) + {\cal F}_B(m)\nonumber\\
&\ge&  \sigma_d \left[1 - {\cal O}\left(\frac{1}{R-2}\right)\right][R-2]_+^{d-1}\left(1 -\kappa \right)^2 S +
\frac{L^d}{ 2\chi_-}(\delta  - 2m_\beta\frac{\sigma_d}{d}(R/L)^d)^2 - \frac{L^d}{ 2\chi_-}\e\ .\nonumber\\
\end{eqnarray}
It now remains to optimize this over $R$.  We note that the lower bound on $\F_B(m)$ decreases as $R$
increases until $R$ is of order $r_0$.  But for such values of $r$, the expression for the lower bound on
the surface contribution simplifies:
$$\left[1 - {\cal O}\left(\frac{1}{R-2}\right)\right][R-2]_+^{d-1} = R^{d-1} + {\cal O}(r_0^{d-2})\ .$$

Now introduce $S_- = S\displaystyle{\left(1 -\frac{\kappa}{m_\beta} \right)^2}$ and $\eta = R^d/r_0^d$. Then we can rewrite this lower bound as
\begin{equation}\label{goodform}
{\cal F}(m)  \ge S_-\sigma_d r_0^{d-1}\left(\eta^{1-1/d} + \frac{S\chi}{ S_-\chi_-}C(n)(1-\eta)^2 \right) - 
\frac{L^d}{ 2\chi_-}\e +{\cal O}(r_0^{d-2})\ .
\end{equation}

By Lemma \ref{lb4}, $|A|/L^d = {\cal O}(\delta^2/\kappa^2)$, and by Lemma~\ref{lb4B},  $R = {\cal O}(L^{\frac d{d+1}})$.  Thus, $(R/L)^d= 
{\cal O}(\delta)$.  Therefore,
$$\e ={ \cal O}\left(\frac{\delta^3}{ \kappa^2} + \delta^3\right)\ .$$
With the choice $\kappa = \delta^{1/3}$, this gives us
$\e = {\cal O}(\delta^{7/3})$, as noted in the remark following  Lemma \ref{lb5}.
The essential point is that this is negligible compared to $\delta^2$ as $L$ tends to infinity in the critical scaling regime.

As $L\to \infty$ in the critical scaling regime, $S_- \to S$ and  $\chi_- \to \chi$. Moreover, from (\ref{deltaequi}) and the definition of $\e$,
$L^d\e/\sigma_d r_0^{d-1} = L^{-\frac{d}{3(d+1)}}\to 0$ as $L\to \infty$. Thus, (\ref{goodform}) provides the lower bound
needed to prove (\ref{limone}). The upper bound is provided by Lemma \ref{uplem}.  The remaining statements
follow from the analysis of the minimization of the phenomenological free energy function $\Phi(\eta)$ that was explained in Section 2 \qed

\subsection{Proof of Theorem \ref{thm2}} %\label{lb1}

Suppose that $n = -m_\beta + KL^{d/(d+1)}$ where $K < K_\star$. We shall show that, in this case, 
 any non-constant trial function $m$ has a higher free energy than the uniform trial function $m(x) = n$, at least for all sufficiently large $L$.

Recalling that $S\sigma_d r_0^{d-1}C(D_0,L) = {\cal F}(n)$, define $\bar\eta$ by
$$\bar\eta = \sup\left\{ \eta \ :\ 
S_-\sigma_d r_0^{d-1}\left(\eta^{1-1/d} + \frac{S\chi}{ S_-\chi_-}C(D_0,L)(1-\eta)^2 \right) - 
\frac{L^d}{ 2\chi_-}\e <  S\sigma_d r_0^{d-1}C(D_0,L) \ \right\}$$

As in the proof of Theorem \ref{thm1}, for all $L$ sufficiently large, $S_-$ is sufficiently close to $S$, and $\chi_-$ is sufficiently close to $\chi$ that 
$$\frac{S\chi }{ S_-\chi_-}C(D_0,L) < C$$
for some $C < C_\star$. For $C < C_\star$, the unique minimizer of 
$$\eta \mapsto \eta^{1-1/d} + C(1-\eta)^2$$
is $\eta =0$. Therefore, since $\e L^d/\sigma_d r_0^{d-1}\to 0$ as $L\to \infty$, it follows that
$\bar\eta \to 0$ as $L\to \infty$. 

Now, as in the previous section, for any non-uniform minimizer $m$, there is a relation between $\eta$
and the size of the level set  $|\{m > m_\beta- \kappa\}|$:   for a given
$\eta = (R/r_0)^{1/d}$,
$|\{m > m_\beta- \kappa\}| = (\sigma_d/d)R^d$.  Here, as in the last section, $\kappa = \delta^{1/3}$ 
with $\delta$ given by (\ref{deltaequi}).  It follows from (\ref{goodform}) and the definition of $\bar\eta$
that for any non-constant minimizer $m$, $\eta < \bar\eta$, and so
$|\{m > m_\beta- \kappa\}|$ is negligibly small compared with $D_0 = (\sigma_d/d)r_0^d$,
the volume of the equimolar ball, when $L$ is large.

In other words, if    $n = -m_\beta + KL^{d/(d+1)}$ where $K < K_\star$, and $L$ is large, then any droplet in any minimizer must be extremely small. To prove Theorem \ref{thm2}, it therefore suffices to show that such extremely small drops are impossible in a minimizing $m$.  The following Lemma gives the required lower bound, and completes the proof. \qed

\medskip
\begin{lm}\label{uni1}  For all $K >0$, there is a constant $c_K>0$ depending only on $K$ so that if $n \le -m_\beta + KL^{-\frac d{d+1}}$
and  $m$ is any non-uniform minimizer for (\ref{GLP}), then
$$\frac{\sigma_d}{d}R^d := |\{x\in {\cal T}_L\, |\, m(x) > m_\beta- \kappa\}| \ge c_K r_0^d\ .$$
Moreover, $c_K$ is uniformly strictly positive for all $K$ in an interval around $K_\star$.
\end{lm}
\medskip

We first explain the idea behind the proof. It is convenient to use (\ref{FtoG})
to write $\F(m)$ in terms of ${\cal G}(\omega)$ as in the proof of Lemma~\ref{lb2}.

We know from Lemma~\ref{lb2}, using the notation from there,  that if $\F(m) \le \F(n)$, then
${\cal G}(\omega) \le 0$.  Since for some $c < \infty$, $G(\omega(x)) > - c\delta$ for all  $x$, and since the
set on which $G(\omega(x))< 0$ is contained in $C$,
\begin{equation}\label{end45}
{\cal G}(\omega) \ge \left[\frac{1}{4}
\int_{A}\int_{A}\dd y  |\omega(x) - \omega(y)|^2J(x-y)\dd x  \dd y  + \int_{A}G(\omega(x))\dd x\right]  - c\delta R^d\ .
\end{equation}

Now, if we could show that for some $\widetilde S > 0$,  
$$
\left[\frac{1}{4}
\int_{A}\int_{A}  |\omega(x) - \omega(y)|^2J(x-y)\dd x  \dd y  + \int_{A}G(\omega(x))\dd x\right] \ge
\frac{\sigma_d}{d}\widetilde S R^{d-1}
\ ,
$$
we would have
$$
{\cal G}(\omega) \ge  \frac{\sigma_d}{d}\widetilde S R^{d-1} -  c\delta R^d = R^{d-1}\left(\frac{\sigma_d}{d}\widetilde S - c\delta R\right)\ ,
$$
and this is strictly positive unless $R$ is on the order of $\delta^{-1}$, 
and in the critical scaling regime, $r_0$ is proportional to $\delta^{-1}$.

Of course
\begin{multline}\label{firsttry}
\frac{1}{4}
\int_{A}\int_{A}  |\omega(x) - \omega(y)|^2J(x-y)\dd x  \dd y  + \int_{A}G(\omega(x))\dd x = \\
\frac{1}{4}
\int_{A}\int_{A} |m(x) - m(y)|^2J(x-y)\dd x  \dd y +  \int_{A}G(m(x)- n)\dd x\ .
\end{multline}
so we can hope to bring the methods of Section 4.3 to bear on this problem. There are two obstacles: 
The first obstacle is that the possible penalty for ``unwrapping the torus'' that is estimated in Lemma~\ref{lbA} is of order $(|A| + |C|)/L$. When $R$ is small, so is $|C|$, but the only
upper bound that we have on $|A|$ is the one provided by Lemma~\ref{lb4}. This is independent of $R$, and so 
it very well can be that for small $R$, $|A|/L$ is large compared to $R^{d-1}$, so that the 
penalty for ``unwrapping the torus'' completely swallows up the surface term.

The second  obstacle is that  lower bound on the surface contribution that we obtained in Lemma~\ref{lbA1} becomes trivial for $R < 2$, twice the range of $J$.

 To deal with the first obstacle, use the fact that on $A$, $G(m(x)- n) \ge c\kappa^2$ for some $c>0$. 
 Therefore,
$$ \int_{A}G(m(x)- n)\dd x = \frac{1}{2} \int_{A}G(m(x)- n)\dd x + \frac{1}{2} |A| c\kappa^2\ .$$
We then have from (\ref{firsttry}) that
\begin{multline}\label{second}
\frac{1}{4}
\int_{A}\int_{A}  |\omega(x) - \omega(y)|^2J(x-y)\dd x  \dd y  + \int_{A}G(\omega(x))\dd x = \\
\frac{1}{4}
\int_{A}\int_{A} |m(x) - m(y)|^2J(x-y)\dd x  \dd y + \frac{1}{2} \int_{A}G(m(x)- n)\dd x +  \frac{1}{2} |A| c\kappa^2 \ .
\end{multline}
Since 
$$\frac{1}{2} |A| c\kappa^2 > \frac{|A|}{L}$$
for all large $L$, the final term in (\ref{second}) more than compensates for the price of ``unwrapping the torus''.

It remains to deal with the second obstacle, as concerns
$$
\frac{1}{4}
\int_{A}\int_{A} |m(x) - m(y)|^2J(x-y)\dd x  \dd y + \frac{1}{2} \int_{A}G(m(x)- n)\dd x\ .
$$

Toward this end, we  first give a simple, {\em  direct} argument, independent of the reasoning in Section 4.3 to show that if $\F(m) \le \F(n)$, and $m$ is not constant, then $R$ is bounded below by a constant of order one. 
Specifically, we shall show that in this case , $R \ge 2^{-(1+1/d)}$.

We then ``cut down'' the range of $J$ so this it is small compared with  $2^{-(1+1/d)}$.  That is,
 for $\rho>0$, define $J_\rho(s) = J(s)$ for $0 \le s < \rho$, and $J_\rho(s) = 0$
 otherwise. Having reduced the range of $J$, the error term in (\ref{curve}) becomes
 ${\displaystyle\left( 1 - {\cal O}\left(\frac{\rho}{R - 2\rho}\right)\right) }$
 for $|x|> R- 2\rho$. We choose $r$ sufficiently  small that this factor is at least $2/3$ for 
 $R \ge 2^{-(1+1/d)}$. This shall take care of the second obstacle. We now provide the detials.

\noindent{\bf Proof of Lemma~\ref{uni1}:} We shall first show that if $\F(m) \le \F(n)$, and $m$ is not constant,
then $R \ge 2^{-(1+1/d)}$. First of all, by  Lemma~\ref{lb2}, if  $\F(m) \le \F(n)$, and $m$ is not constant, then 
$C$ is not empty, and at least we know $R>0$. Let us improve on this.

With $\omega_-$ defined as in  Lemma~\ref{lb2}, define
${\displaystyle\widetilde C := \{\ x\ :\ \omega(x) \ge \omega_-\ \}}$,
and define $\widetilde R$ so that ${\displaystyle \frac{\sigma_d}{d}\widetilde R^d = |\widetilde C|}$.
Then by Lemma~\ref{lb2}, there is a constant $c_1$ so that 
${\displaystyle \int_{{\cal T}_L} G(\omega(x))\dd x \ge  -c_1\delta \widetilde R^d}$.

We now claim that if $R < 2^{-(1+1/d)}$, the interaction term makes a much larger positive contribution 
to ${\cal G}(\omega)$, so that  ${\cal G}(\omega)>0$, which would imply $\F(m) > \F(n)$.

To see this, note that by Lemma~\ref{lb2}, if $x\in \widetilde C$, and $y\notin C$, then for some positive constant $c$,
$$|m(x) - m(y)|^2 = |\omega(x) - \omega(y)|^2 \ge c\kappa^2 =  c\delta^{2/3}\ ,$$
since $m(y) \le m_\beta - \kappa$, but $m(x) \ge m_\beta - {\cal O}(\delta^{1/2})$.

Now, by our assumptions on $J$,  for each $x\in \widetilde C$, $J(x-y) \ge a>0$ on the ball  of radius $1/2$ about $x$.  If   $R < 2^{-(1+1/d)}$, then $C$ can fill up at most one half of this ball, and so the volume of the set of
points $y$ in  the ball  of radius $1/2$ about $x$ for which $y\notin C$ is at least ${\displaystyle
\frac{\sigma_d}{d}2^{-(d+1)}}$. Hence
$$\int_{{\cal T}_L}\dd x \int_{{\cal T}_L}\dd y  |\omega(x) - \omega(y)|^2 \ge
\int_{\widetilde C}\dd x \int_{C^c}\dd y  |\omega(x) - \omega(y)|^2 J(x-y)\ge   c_2\delta^{2/3} \widetilde R^d\ ,$$
for some $c_2>0$.  Altogether,
${\displaystyle{\cal G}(\omega) \ge [c_2\delta^{2/3} - c_1\delta]\widetilde R^d}$,
and this is strictly positive for $L$ large enough. 
Therefore, we may assume without loss of generality that $m$ is such that  $R > 2^{-(1+1/d)}$.

We return to (\ref{second}), and ``cut down'' the range of $J$, replacing $J$ by $J_\rho$, $\rho < 1$, which clearly decreases the right hand side of (\ref{second})
\begin{multline}\label{third}
\frac{1}{4}
\int_{A}\int_{A} |m(x) - m(y)|^2J(x-y)\dd x  \dd y + \frac{1}{2} \int_{A}G(m(x)- n)\dd x +  \frac{1}{2} |A| c\kappa^2 \ge \\
\frac{1}{4}
\int_{A}\int_{A} |m(x) - m(y)|^2J_\rho(x-y)\dd x  \dd y + \frac{1}{2} \int_{A}G(m(x)- n)\dd x +  \frac{1}{2} |A| c\kappa^2 \ .
\end{multline}

Next, by (\ref{trunc2}) and Lemma~\ref{lbA}, we have
\begin{equation}\label{fourth}
\int_{A}\int_{A} |m(x) - m(y)|^2J_\rho(x-y)\dd x  \dd y \ge 
\int_{\R^d}\int_{\R^d} |\widetilde m(x) - \widetilde m(y)|^2J_\rho(x-y)\dd x  \dd y - 2d\frac{|A|+|C|}{L}\ .
\end{equation}

Now let  $\overline J_\rho$ be
 defined in terms of $J_\rho$ just as $\overline J$ is defined in terms of $J$, and let 
 $m^*$ denote the  the rearrangement
 of $\widetilde m$, as before.   Having reduced the range of $J$ from $1$ to $\rho$, the estimate (\ref{tou333}) becomes
 \begin{multline}\label{tou400}
\int_{\R^d}  \int_{\R^d}  | m^*(x) -  m^*(y)|^2 J_\rho (x-y) \dd x\dd y =\\
\left[1- {\cal O}\left(\frac{\rho}{R-2\rho}\right)\right]_+ [R-\rho]_+\int_{\R} \int_{\R}  | m^*(r) -  m^*(s)|^2\overline J_\rho (r-s)\sigma_d \dd r\dd s\ ,\end{multline}
where once again, $[\cdot]_+$ is the positive part function.

We now choose $0<\rho<2^{-(2+1/d)}$ small enough that the prefactor ${\displaystyle \left[1- {\cal O}\left(\frac{\rho}{R-2\rho}\right)\right]_+ [R-\rho]_+}$ is at least ${\displaystyle \frac{1}{2}R^{d-1}}$ for $R \ge 2^{-(1+1/d)}$. 
Of course, we also have
\begin{equation}\label{end4}
\int_{A}G(\omega(x))\dd x \ge \sigma_d R^{d-1}\int_{\R}\widetilde G(m^*(z)- n)\dd z
\end{equation}
where as in (\ref{tou9}), we define
$\widetilde G$ by
\begin{equation}\label{tou999} 
\widetilde G(\omega) = 
\begin{cases} 
G(\omega)  &{\rm  if}\quad  h_+> \omega+n > h_- \\ 
0  &{\rm  if}\quad  \omega+n \le h_- \ \text{ or} \ m\ge h_+\\ 
\end{cases}\ .
\end{equation}

Combining (\ref{third}),  (\ref{fourth}), (\ref{tou400}) and (\ref{end4}), we  have
\begin{multline}\label{sixth}
\frac{1}{4}
\int_{A}\int_{A} |m(x) - m(y)|^2J(x-y)\dd x  \dd y + \frac{1}{2} \int_{A}G(m(x)- n)\dd x +  \frac{1}{2} |A| c\kappa^2 \ge \\
R^{d-1}\sigma_d\left[
\frac{1}{8}
\int_{\R}\int_{\R} |m(r) - m(r)|^2\overline J_\rho(x-y)\sigma_d \dd r  \dd s + \frac{1}{2} \int_{A}\widetilde
G(m(r)- n)\dd r\right]  - d\frac{|A|+|C|}{2L} +  \frac{1}{2} |A| c\kappa^2 \ .
\end{multline}

Now let $\widetilde S$ be defined by
\begin{equation}
\widetilde S  = \inf\left\{\ \frac{1}{2}\int_{\R}  {\rm d}z \left( G(m^*(z)- n)
+\frac{1}{ 4}\int_{\R}|m(z) - m(z')|^2\bar J_\rho(z-z'){\rm d}z' \right)\ :\ 
\lim_{z\to \pm \infty}m(z) = h_{\mp} \ \right\}\ \label{pst2}
\end{equation}

Combining (\ref{end45})  (\ref{second})   and (\ref{sixth}), we finally have
$$ {\cal G}(\omega) \ge   \widetilde S \frac{\sigma_d}{d}R^{d-1}  - d\frac{|A| + (\sigma_d/d)R^d}{2L} + \frac{1}{2}c\kappa^2 |A| \ .$$
 Since $\kappa^2 = L^{-\frac{2d}{3(d+1)}}$, for large $L$, 
 ${\displaystyle-2\frac{|A| }{2L} + \frac{1}{2}c\kappa^2 |A| > 0}$, and
  $${\cal G}(\omega) \ge   \widetilde S \frac{\sigma_d}{d}R^{d-1}  - c\frac{\sigma_dR^d}{dL} - c\delta R^d
 = R^{d-1}\left( \widetilde S \frac{\sigma_d}{d}  - c\frac{\sigma_dR}{dL} - c\delta R\right)\ .$$
 This is positive unless $R> c\delta^{-1}$, and by (\ref{FtoG}) we conclude that whenever $\F(m) \le \F(n)$,
and $m$ is not constant,  $R  > c\delta^{-1}$.
 Since in the critical scaling regime, $r_0$ is proportional to $\delta^{-1}$, this proves the lemma. \qed

\medskip

\subsection{Proof of Theorem \ref{thm3}.} Suppose $m$ is such that ${\cal F}(m)\le f_L(n)+\alpha$, for some given $\alpha>0$. With the notation of the previous sections, let $\eta=|C|/D_0$ and $\eta_c$ be the optimal volume fraction corresponding to $n$. Then, by (\ref{goodform}) we get
\begin{eqnarray}
{\cal F}(m)&\ge& S_-\big[\Phi(\eta)-\Phi(\eta_c)\big]+S_-\Phi(\eta_c)+ o(L^{\frac d{d+1}})\\&\ge&  S_-\big[\Phi(\eta)-\Phi(\eta_c)\big]+f_L(n)+o(L^{\frac d{d+1}}).\nonumber
\end{eqnarray}
This entails that 
$\Phi(\eta)-\Phi(\eta_c)\le \frac{\alpha}{S_-}+o(L^{\frac d{d+1}})$.
By the definition of $\Phi$, there is a constant $\varphi_0$ such that
${\displaystyle \varphi_0(\eta-\eta_c)^2\le {\alpha}{L^{-\frac d{d+1}}}+o(1)}$,
so, for $L$ sufficiently large
${\displaystyle \left|\frac{|C|}{D_0}-\eta\right|\le o(1)}$.\qed

\section{The shape problem}

%\subsection{The shape problem}

Throughout this section, we use the notation defined in (\ref{hpmdef}) and 
(\ref{ABCdef}). In particular, given a trial function $m$, the set $C$ is the set of points $x$ on which $m(x)$ takes values that are
close to $m_\beta$ or larger, $B$ is the set of points $x$ on which $m(x)$ takes values that are
close to $-m_\beta$ or smaller, and $A$ is everything else. We shall also assume that $m$ is 
defined on all of $\R^d$; we have already seen how to extend $m$ from ${\cal T}_L$ to $\R^d$ with negligible cost in free energy, so let us suppose this is done, and $|A|$ and $|C|$  are finite. Finally, in this subsection we use the notation (\ref{Rdef}), so that 
$R$ is the radius of the ball in $\R^d$
with the same Lebesgue measure as $C$.  Up to now, we have been concerned with the sizes of $A$ and $C$.
Going forward,  we are concerned with their shape. It is easier to get control of this in the local case.

In the case of the local (Allen-Cahn or van der Waals) free energy functional (\ref{AC}), the lower bound (\ref{coco})
brings the surface area of the boundary of $C$ into the lower bound on the free energy. Then stability results
for the isoperimetric inequality \cite{Bo, H, HHW} can be used to show \cite{CCELM1}  that if $C$ is not nearly spherical, there is a significant cost in free energy.  It is an open problem, which we refer to as the {\em shape problem},
to prove this for the  Gates-Lebowitz-Penrose free energy functional.
To clarify the difference between the local and non-local cases, let us briefly recall an argument from \cite{CCELM1}. 
(We shall in fact improve the result in \cite{CCELM1} by using a new stability inequality from \cite{FMP}.)

Let $E$ be a Borel measurable set in $\R^d$. As usual, let $|E|$ denote is Lebesgue measure, and let $P(E)$
denote its {\em perimeter}.  If the set $E$ is sufficiently regular, $P(E)$ is the $d-1$ dimensional Haussdorf measure of the boundary of $E$, though for very irregular sets it can be much smaller. The perimeter functional is an extension of the surface area functional to general Borel sets, due to De Giorgi \cite{DG}  , with the key property  that it enjoys good lower semicontinuity properties. See Maggi's review \cite{Ma} for more information. 

The {\em isoperimetric deficit} of $E$, $\delta(E)$,  is the quantity
\begin{equation}\label{def}
\delta(E) = \frac{P(E)}
{ d^{(d-1)/d} \sigma_d^{1/d} |E|^{(d-1)/d} }-1\ .
\end{equation}
The general isoperimetric inequality of De Giorgi says that $\delta(E)\ge 0$ with equality if and only if $E$, up to a set of measure zero, is a ball. 

The stability results that we refer to give a lower bound on $\delta(E)$ in terms of the {\em Fraenkel asymmetry} 
of $E$, $A(E)$, which measures the extent to which $E$ differs  from being a ball:
\begin{equation}\label{frae}
A(E) = \inf\left\{\ \frac{|E \Delta B(r,x)|}{|E|}\ :\ \frac{\sigma_d}{d}r^d = |E|\ , \ x\in \R^d\ \right\}
\end{equation}
where $B(r,x)$ is the ball of radius $r$ centered on $x$ in $\R^d$, and $E\Delta B$ denotes the 
{\em symmetric difference}
of $E$ and $B$; that is,
$$E\Delta B = (E\backslash B)\cup (B\backslash E)\ .$$

The theorem of Fusco, Maggi and Pratelli \cite{FMP} says that for all $d\ge 2$, there is a constant $C(d)$
depending only on $d$, so that for all Borel sets $E$ in $\R^d$,
\begin{equation}\label{mdef}
P(E) \ge [1+ C(d)A^2(E)]\sigma_d\left(\frac{d}{\sigma_d}|E|\right)^{(d-1)/d}\ .
\end{equation}
This improved on an earlier result  of Hall, Hayman and Weitsman \cite{H,HHW} of the same character,
but with $A^4(E)$ in place of $A^2(E)$. The exponent $2$ is sharp; see \cite{Ma} for further discussion
and background.

To apply this result to the shape problem for the local free energy functional,  suppose
 that for some trial function $m$ and some
$\epsilon,\eta >0$, one has that the volume of $C$ lying outside {\em every} ball of radius $(1+\epsilon)R$
is at least $\eta |C|$.  

Now let $E_h$ be defined by
$$E_h :=  \{\ x \ :\ m(x) \ge h\ \} \ ,$$
so that $E_{h_+} = C$, and  $E_{h_-} = A\cup C$. By containment, if $k > h$, then $E_k \subset E_h$. 
This  has the
consequence that for all $h\in [h_-,h_+]$, the 
volume of $E_h$ lying outside {\em every} ball of radius $(1+\epsilon)R$
is at least $\eta |C|$.   

Moreover, since in the critical scaling regime, Lemma~\ref{lb4} says that $|A|$ is negligible compared to $|C|$, for all sufficiently large $L$, and all $h\in [h_-,h_+]$,  the radius of a ball in $\R^d$ with the same volume as $E_h$
is no greater than $(1+\epsilon)R$.  

Consequently, for each $h\in [h_-,h_+]$,
$A(E_h) \ge \eta$. Therefore, by (\ref{mdef}), and one more use of $|E_h| \ge |E_{h_+}| = |C|$, 
$$
P(E_h) \ge (1+ C(d)\eta^2)\sigma_d R^{(d-1)/d}\ .
$$
In the context of (\ref{coco}), for almost every $h$, $P(E_h) = |\Gamma_h|$,  the $d-1$ dimensional Haussdorf measure of
the set $\Gamma_h = \{\ x\ :\ m(x) = h\ \}$, and hence we have a lower bound on $|\Gamma_h|$ that is uniform in
$h$. Using this in (\ref{coco}), 
 one gets a lower bound  on the free energy of $m$  that is larger than the minimizing value
by a factor of (essentially) $(1+ C(d)\eta^2)$. Hence if the free energy of a trial function is sufficiently
close to the minimizing value, $\eta$ must be correspondingly small. This forces $C$ to be very nearly a ball.
A similar argument applies to each of the $E_h$ for  $h\in (h_-,h_+)$, although the degree of control on the roundness of $E_h$ diminishes as $h$ approaches $h_-$.  A precise
statement in terms of the $L^p$ distance between $m$ and an ideal round droplet profile  may be found in \cite{CCELM1}.

It is an interesting open problem to develop bounds of this type that would apply to the Gates-Lebowitz-Penrose free energy functional. It would be very surprising if the physical model behind it did not capture enough physical reality to
control the shape of droplets in near-minimizers.   We note that one can apply stability for the isoperimetric inequality to the GLP functional, but only in the {\em sharp interface scaling limit}, as discussed in \cite{P}.
However the size of our critical droplet goes to zero in this limit.  

What would seem to be useful here would be a stability result for the Riesz rearrangement inequality, or at least the special case in which one of the three functions is already
rearranged, and has all centered balls of sufficiently small radius as level sets. 
A general stability result for it might be quite subtle; the cases of equality were only determined relatively recently by Burchard \cite{Bu}. However, for $J$ is radially symmetric and well-behaved, there is a simpler and cleaner result on the cases of equality due to Lieb \cite{L}, and in this setting, one might expect a stability result
that would force the other two functions to be nearly rearranged for near minimizers. 
This will be the subject of future research.

\medskip
\centerline{\bf Acknowledgments}
\medskip

We would like to thank Thierry Bodineau and Errico Presutti  for very helpful discussions.
E. C. was partially supported by  U.S. NSF grant DMS 06-00037. M.C.C. was partially supported by unda‹o para a Cincia e a
Tecnologia, Financiamento Base 2008 - ISFL/1/209 and
grant POCI/MAT/61931/2004. R.M. and R. E. were partially supported by MIUR, INDAM-GNFM. J.L.L. was supported by AFORS grant  AF
49620-01-1-0154 and  NSF Grant DMR-044-2066.
This work has also been supported by the European Commission through its 6th Framework Program
``Structuring the European Research Area'' and the contract Nr. RITA-CT-2004-505493 for the provision of Transnational Access implemented
as Specific Support Action.

\end{document}